\documentclass[aps,pra,twocolumn,showpacs,floatfix,footinbib,groupedaddress,superscriptaddress]{revtex4}
\usepackage{amsmath}
\usepackage{amssymb,graphics,graphicx,times,bm,bbm,multirow}
\usepackage{footnote}
\usepackage{appendix}

\begin{document}

\title{Efficient estimation of energy transfer efficiency in
light-harvesting complexes}
\author{A. Shabani}
\affiliation{Department of Chemistry, Princeton University, Princeton, New
Jersey 08544}
\author{M. Mohseni}
\affiliation{Center for Excitonics, Research Laboratory of Electronics,
Massachusetts Institute of Technology, Cambridge, MA 02139}
\author{H. Rabitz}
\affiliation{Department of Chemistry, Princeton University, Princeton, New
Jersey 08544}
\author{S. Lloyd}
\affiliation{Department of Mechanical Engineering, Massachusetts Institute of
Technology,
Cambridge, MA 02139}

\begin{abstract}
The fundamental physical mechanisms of energy transfer in
photosynthetic complexes is not yet fully understood. In particular,
the degree of efficiency or sensitivity of these systems for energy
transfer is not known given their non-perturbative and non-Markovian
interactions with proteins backbone and surrounding photonic and
phononic environments. One major problem in studying
light-harvesting complexes has been the lack of an efficient method
for simulation of their dynamics in biological environments. To this
end, here we revisit the second-order time-convolution (TC2) master
equation and examine its reliability beyond extreme Markovian and
perturbative limits. In particular, we present a derivation of TC2 without making the 
usual weak system-bath coupling assumption. Using this equation, we explore the long time behaviour of exciton dynamics of Fenna-Matthews-Olson (FMO) portein complex.
Moreover, we introduce a constructive
error analysis to estimate the accuracy of TC2 equation in
calculating energy transfer efficiency, exhibiting reliable
performance for environments with weak and intermediate memory and
strength. Furthermore, we numerically show that energy transfer efficiency is optimal
and robust for the FMO protein complex of green sulphur bacteria
with respect to variations in reorganization energy and bath
correlation time-scales.
\end{abstract}

\maketitle

\section{Introduction}

Over the past few decades, there have has been significant interests in
monitoring and simulating excitonic energy transfer in molecular systems
\cite{Mukamel:Book, lit1,lit2,lit3,lit4,may1,lit5,may2,lit6,lit7,Adolph,lit8}.
Recently 2D
electronic spectroscopy demonstrated that the excitation energy transfer in
photosynthetic complexes could involve long-lived quantum
coherence \cite{Engel07,Lee07,Calhoun,Mercer09,Scholes09-1,Scholes09-2,panit}. These experimental observations have
lead to vigorous theoretical efforts to study quantum coherent
dynamics in light-harvesting complexes
\cite{mohseni-fmo,Plenio08-1,Rebentrost08-1,Rebentrost08-2,Castro08,Plenio09,
AkiPNAS, Asadian,Aki-PCCP,Shim12} and observations of environment-assisted quantum
transport \cite{mohseni-fmo,Plenio08-1,Rebentrost08-2}. Moreover, various ways for partitioning the
contribution of quantum coherence to the energy transfer efficiency
(ETE) have been explored
\cite{Rebentrost08-1,Plenio09,CaoSilbey,Sarovar,Fassioli,Plenio10}.
In spite of many years of theoretical studies in energy transfer,
the role of quantum effects in the
biological performance of photosynthetic systems is not fully
understood. In particular, it is not known whether it is necessary
to include quantum effects to demonstrate the optimal efficiency of
these systems, predict the outcomes of ultrafast spectroscopic
experiments \cite{masoud-tomography}, and explain the evolutionary
path of the photosynthesis complexes \cite{Xiong}.

The major difficulty in studying such complex open quantum system
dynamics arises from the lack of an efficient method for simulation
under realistic conditions. In the relevant biological systems, the
system-bath coupling strength and free Hamiltonian parameters have
typically comparable strength. In such cases the popular methods
developed for the extreme perturbative regimes of weak system or
weak environment break down, such as F\"{o}rster energy transfer or
Redfield/Lindblad formalisms \cite{Ishizaki09}.
Moreover, the stochastic master equation approaches lead to an incomplete
description of the dynamics, for instance the Haken-Strobel-Reineker model
\cite{H-S,H-S-1},
treating environment as a classical white noise, is unable to
capture finite temperature limit and detailed bath spectral density
\cite{Ishizaki09}. Additionally, the
bath has often enough memory that all the master equation methods
based on Markovian assumption become inadequate. A hierarchy of
coupled master equations has been recently presented by Ishizaki and
Fleming \cite{Ishizaki09-2,AkiPNAS}, based on earlier works of Kubo
and Tanimura \cite{Tanimura,Tanimura-1,Tanimura2}, that provides a general benchmark for
simulation of light-harvesting complexes in all non-perturbative and
non-markovian regimes with \emph{arbitrary accuracy}. However, these
general methods inevitably involve significant computational
overhead with increasing the size of system, temporal coherence, and
in the low temperature limit. Thus, a variety of alternative
approaches for simulation of open quantum systems and
light-harvesting complexes have been proposed to capture certain aspects of
non-Markovian
effects, quantum coherence and/or beyond second order perturbation corrections
\cite{Cao,N1,N2,Jang07,N3,Patrick3,N4,N5,Shi,N6,N7,N8,N8-1,N9,N10}.

In this work, we demonstrate that the second-order time-convolution
master equation (TC2) can be applied for approximate calculation of
Energy Transfer Efficiency (ETE) of complex open quantum systems
interacting with a Gaussion environment beyond extreme Markovian and
perturbative limits. We first introduce a set of approximations to
derive the TC2 without the usual weak system-bath coupling
assumption. Our study is based on the earlier work of Cao on the
generalized Bloch-Redfield master equation \cite{Cao}. Similarly, a
post-perturbative derivation of Redfield equations can be obtained
by employing harmonic oscillator baths and the Markov approximation
\cite{Aki-PCCP}. It has also been recently shown that, for dichotomic
random two-jump process, a Stochastic Liouville equations can
capture the excitonic dynamics beyond the limits of both the weak
coupling and short time correlation conditions \cite{Bourret-1}. A main task
is to quantify how well such methods can capture dynamical
properties of excitonic system in the intermediate regimes. Here, we
provide a combination of analytical and phenomenological approaches
for estimating errors, due to ignoring dynamical contribution of
certain higher-order bath correlation functions, for calculation of
ETE via TC2 master equation. This allows us to quantify the regions
that TC2 would breaks downs for such computation in the presence of
environments with strong memory and strength. Our numerical
simulations demonstrate that the estimated values of system-bath
coupling strength and bath memory time-scale for the FMO protein
indeed lead to optimal and robust energy transport in the
intermediate regimes. This analysis could allow for a practical way
to quantify the performance of large light-harvesting complexes and
to explore the optimality and robustness of these systems by relying
on a single time-convolution master equation.

In a companion manuscript \cite{comp}, we use this technique to
comprehensively explore the efficiency of FMO protein and other
small-size light-harvesting geometries. We demonstrate their
optimality and robustness with respect to all the relevant internal
and environmental parameters including, multichromophores spatial
compactness, number of chromophores, spatial connectivity, dipole moments
orientations,
disorders, excitonic band gap structures, reorganization energy,
temperature, bath spatial and temporal correlations, initial
excitations, and trapping mechanisms. We address whether or not the
FMO complex structure and typically non-perturbative and
non-Markovian environmental interactions are necessary for its
performance. Specifically, we explore the general design principles
for achieving optimal and robust excitonic energy transport and
whether there are fundamental reasons for the convergence of
time scales with resect to internal parameters, environmental
interactions, and trapping mechanisms.

This article is organized as follows. Section II discusses general energy
transport dynamics and its modeling in complex quantum systems. In section III, the
definition of ETE is introduced and for FMO complex, the optimality and robustness of ETE as a
function of system-bath coupling
and bath memory is presented. In section IV, we evaluate the accuracy of our
energy transport simulation. Detailed mathematical derivation of the
TC2 equation and the error analysis for the approximate estimation of ETE are
presented in the appendices.

\section{Exciton transport in complex quantum systems}

Excitons are quasiparticles, each formed from a pair of electron and
hole, that provide a natural means to convert energy between photons
and electrons. There are two well-studied theoretical regimes of
exciton transport. One extreme limit is semiclassical F\"{o}rster
theory in which the electronic Coulomb interaction among different
chromophores is treated perturbatively compared to the typically
strong electron-phonon coupling. This condition leads to incoherent
classical walks of exciton hopping among the chromophores. In the
other extreme limit, the electron-phonon interaction is
treated perturbatively using Redfield or Lindblad dynamical equations \cite%
{mohseni-fmo}. However, the biologically relevant but less-studied cases are
within the intermediate regime when the strength of the Coulomb and
electron-phonon interactions are comparable. The HEOM in
principle can be applied to all of regimes \cite{AkiPNAS,Ishizaki09,Tanimura,Tanimura-1,Tanimura2}.
However, this approach is inefficient for exploring the
properties of large light-harvesting complexes over a wide range of
parameters, and for other applications such as optimal material design \cite{thebook}.
\begin{figure}[tp]
\includegraphics[width=8cm,height=6cm]{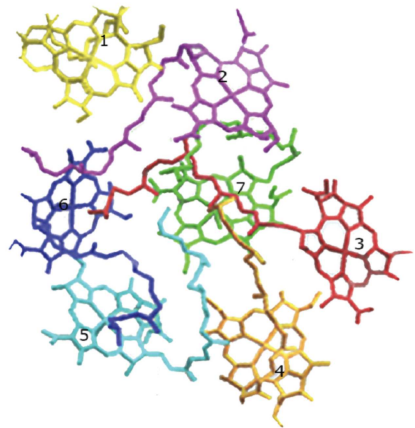}
\caption{The disordered structure of the Fenna-Matthews-Olson (FMO)
complex: It is a trimer consisting of three identical monomers each
formed by seven Bacteriochlorophylls (BChl) embedded in a
scaffold protein. The FMO complex acts as an energy transfer channel in
green sulphur bacteria guiding excitons from the light-harvesting
antenna complex, in the proximity of BChls 1 and 6, to the reaction
center which is in the proximity of BChls 3 and 4}. \label{fmo}
\end{figure}

Here, in order to arrive at a more efficient but less accurate approach, we start by considering the general
time-evolution of open quantum systems. The dynamics of a photosynthetic system
is influenced by the surrounding scaffold protein and solvent. Such an
environment can be modeled as a phonon bath consisting of a set of harmonic
oscillators. The total system-bath Hamiltonian can be written as
\begin{equation}
H_{total}=H_{S}+H_{ph}+H_{S-ph}\label{Hamiltonian}
\end{equation}%
where
\begin{eqnarray}
H_{S} &=&\sum_{j,k}\epsilon _{j}|j\rangle \langle j|+J_{j,k}|j\rangle
\langle k|, \notag \\
H_{ph} &=&\sum_{j,\xi }(\frac{p_{j,\xi }^{2}}{2m_\xi}+\frac{m_\xi\omega_\xi^2 q_{j,\xi
}^{2}}{2}), \notag \\
H_{S-ph} &=&\sum_{j}S_{j}B_{j}\label{Hamiltonian2}.
\end{eqnarray}%
Here $|j\rangle $ represents the single exciton state of site $j$. The variables $\omega_\xi$, $p_{j,\xi }$ and $q_{j,\xi }$
are the frequency, position and momentum operators of the oscillator, respectively. The diagonal
elements $\{\epsilon _{j}\}$s include system internal site energies plus
reorganization energy shifts $\lambda _{j}=\sum_{\xi }\hbar \omega _{\xi }d_{j,\xi }^{2}/2$ induced by coupling to the
phonon bath where $d_{j,\xi}$ is the dimensionless displacement
of the $(j,\xi)$th phonon mode from its equilibrium configuration.
 The off-diagonal coefficients $\{J_{j,k}\}$s represent
dipole-dipole interaction between chromophores in different sites. We assume
that each site is linearly interacting with a separate bath, with operators $S_{j}=|j\rangle \langle j|$
and $B_{j}=-\sum_{\xi }\hbar \omega _{\xi }d_{j,\xi }q_{j,\xi }$ accordingly being the system and bath operators.
A more general Hamiltonian should include system-bath coupling terms $\sum_{ij}S_{ij}B_{ij}$ with the system operators $S_{ij}=|i\rangle \langle j|$,
however the fluctuations of the inter-dipole coupling amplitudes are typically much smaller than the site energy fluctuations therefore we ignore
these terms in our analysis \cite{Cho,Prall}.

The dynamics of an open system is given by quantum Liouvillian equation
\begin{equation}
\frac{\partial \rho (t)}{\partial t}=\mathcal{L}_{total}[\rho_{total}
(t)]=-i\hbar \langle \lbrack H_{total},\rho_{total} (t)]\rangle _{ph}\label{Liou}
\end{equation}%
with $\rho_{total}$ denoting the system-bath quantum state,  and $\langle ...\rangle _{ph}$
being an average over phonon bath degrees of freedom. The Liouvillian
superoperator $\mathcal{L}_{total}$ is the sum of superoperators $\mathcal{L}%
_{S}$, $\mathcal{L}_{ph}$ and $\mathcal{L}_{S-ph}$ corresponding to $H_{S}$,
$H_{ph}$ and $H_{S-ph}$. The explicit form of $\mathcal{L}_{total}$ can be
obtained if the system and bath start in a product state: $\rho
_{S-ph}(0)=\rho (0)\otimes \rho _{ph}(0)$. Furthermore, we assume that the phonon
bath is initially in thermal equilibrium state at temperature $T$, $\exp (-\beta H_{ph})/Tr(\exp
(-\beta H_{ph}))$ where $\beta=1/kT$. The assumption of an initial product state can be justified as the
photosystem is in its electronic ground state prior
to interaction with a light source; if a quantum system is
in a pure state, that implying a product system-bath state \cite{Shabani}.

In the interaction picture, the compact formal solution of
Eq. (\ref{Liou}) is
\begin{equation}
\tilde{\rho}(t)=\langle \mathcal{T}_{+}\exp \Big [\int_{0}^{t}\tilde{%
\mathcal{L}}_{total}(s)ds\Big ]\rangle _{ph}\rho(0)\label{formalsolution}
\end{equation}
where $\tilde{O}$ denotes the interaction picture representation for
an operator $O$. Expansion of the above time-ordered exponential
function yields a Dyson expansion for time
evolution of the density operator involving high-order bath correlation
functions. Hereon we drop the subscript $\mathit{total}$ and $\mathit{ph}$ for simplicity.
\begin{align}
\frac{\partial }{\partial t}\tilde{\rho}(t)=\sum_n
\int_{0}^{t}dt_{1}...\int_{0}^{t_{n-1}}dt_{n}
\langle\tilde{\mathcal{L}}(t)\tilde{\mathcal{L}}(t_{1})...\tilde{\mathcal{L}}(t_{n})\rangle
\rho(0)\label{cumul}
\end{align}
 where the $n$-time correlation superoperator has the following form
 \begin{gather}
\langle \tilde{\mathcal{L}}(t_1)...\tilde{\mathcal{L}}(t_n)\rangle\rho(0)=(-i)^n\sum_{j_1..j_n}\sum_{i_1..i_n}(-1)^{n-k}\notag\\
\times\langle \tilde{B}_{j_{k+1}}(t_{i_{k+1}})...\tilde{B}_{j_n}(t_{i_{n}})\tilde{B}_{j_1}(t_{i_{1}})...\tilde{B}_{j_k}(t_{i_k})\rangle \notag\\
\times \tilde{S}_{j_1}(t_{i_1})...\tilde{S}_{j_k}(t_{i_k})\rho(0)\tilde{S}_{j_{k+1}}(t_{i_{k+1}})...\tilde{S}_{j_n}(t_{i_n})\label{nterm}
\end{gather}
with the site indices $\{j_1,...,j_n\}$ and the second summation over all indices
$\{i_1,...,i_n\}\in\{1,..,n\}$ such that the $t_{i_1},...,t_{i_k}$
and $t_{i_{k+1}},...,t_{i_n}$ are ordered backward and forward in
time, respectively. The bath correlation functions vanish in a
finite time interval and the system operator in each term of the
above expansion Eq. (\ref{cumul}) has a bounded norm. Under these
conditions the Dyson expansion always converges \cite{Rugh}.

For the considered system-bath interaction (\ref{Hamiltonian}) the
bath operators $\tilde{B}(t_{j})$ satisfies Gaussian statistics.
That is, $n$-time correlation functions vanishes if $n$ is odd. For
even $n$, according to Gaussian property, the terms up to the second
order in the bath correlation function are sufficient to
\emph{exactly} describe the dynamics of the system:
\begin{eqnarray}
\langle \tilde{B}(t_{i_1})...\tilde{B}%
(t_{i_{2n}})\rangle=\sum_{pairs}\prod_{l,k}\langle \mathcal{I}_{+}\tilde{B}%
(t_{i_k})\tilde{B}(t_{i_{l}})\rangle\label{Wickb}
\end{eqnarray}
where the index {\it{pairs}} denotes the division of the labels $1$ to $2n$ into $n$ unordered pairs.
The operator $\mathcal{I}_{+}$ is the index ordering operator preserving the order of operators
on RHS of Eq.(\ref{Wickb}) similar to the LHS.
Note that here we have applied a generalized Wick's theorem in the form of Wightman functions
rather than the usual form with time-ordered correlation functions \cite{gWick1,gWick2}.
This expansion is possible assuming $\langle \tilde{B}_j(t_1)\tilde{B}_j(t_2)\rangle=\langle \tilde{B}_j(t_2)\tilde{B}_j(t_1)\rangle^*$
in consistence with Kubo-Martin-Schwinger condition \cite{Breuer}.
The most general method to solve the master equation (\ref{Liou}) is to utilize a path
integral formalism leading to HEOM \cite{Tanimura2}. Here, we would like to avoid
such a general approach since the required computational resources grows rapidly
with the increasing size of the system as well as decreasing bath
cut-off frequency and ambient temperature.

In order to obtain a numerically efficient approach for simulation of
complex excitonic systems we first need to understand how the computational
inefficiency arises. It should be noted that the source of computational complexity here
is different from that faced in quantum chemistry and
condensed matter physics in $\it{ab-initio}$ calculations of the ground state
energy of interacting many-body fermionic systems. In such cases, the
Hilbert space grows exponentially with the number of particles and the
degrees of freedom involved. However, efficient techniques such as density
functional theory \cite{DFT} and density matrix renormalization group \cite{DMRG} can provide
approximate solutions.
Here, due to the low intensity incident light
and the time-scale separation of recombination process (in $1$ ns) with fast energy transfer process (in $1$ ps),
we can ignore transitions between multi-excitation levels and focus on energy transport
dynamics in a single exciton manifold. Thus, the
Hilbert space grows linearly with the number of chromophores. However,
due to the open nature of these complexes interacting with an environment that
has strong memory and strength, the time-nonlocal features of the dynamics
are extremely difficult to simulate and explore.
In these cases, reducing the number of environmental
degrees of freedom (e.g, having a smaller bath frequency cutoff) does not translate
into lower computational complexity. On the contrary it will enhance the
non-Markovian behavior of the system. Specifically, the
computational cost raises when attempting to treat these systems
non-perturbatively while mapping the memory effects (e.g. encoded into a
time-nonlocal kernel) to a set of coupled time-local master equations
such as HEOM. In this work, we explore and quantify the applicability of TC2 master equation, as an inherently efficient method to capture certain time non-local feature for calculation of ETE in environments with weak and intermediate strength.   We first present a derivation of TC2 without the usual weak system-bath coupling assumption and quantify the errors in computing ETE as functions of both bath memory and strength.

Here, we outline the main assumptions and steps of our derivation,
for more details see Appendix A. First, we assume that the bath
fluctuations are stationary; that is, these processes are
insensitive to the reference point in time. For such quantum
processes one can express bath correlations
$\langle\tilde{B}_j(t)\tilde{B}_j(t')\rangle$ only as a function of
$t-t'.$ This character has also been assumed in HEOM. The
correlation function is calculated from the bath spectral density
\begin{eqnarray*}
J_j(\omega)=\frac{1}{\hbar}\sum_\xi \frac{d_{j,\xi}^2}{2m_{\xi}\omega_\xi}\delta(\omega-\omega_\xi)
\end{eqnarray*}
as
\begin{eqnarray}
\langle\tilde{B}_j(t)\tilde{B}_j(t')\rangle=\frac{1}{\pi}\int_0^\infty d\omega J_j(\omega)\frac{\exp(-i\omega (t-t'))}{1-\exp(-\beta\hbar\omega)}\label{bathcor}
\end{eqnarray}

As we discuss in Appendix A, the stationary property
of bath fluctuations can be exploited to provide a rather straightforward solution for the equation of
motion in the frequency domain. We obtain this time-nonlocal
master equation by truncating the generalized Wick's expansion (\ref{Wickb}) to the sum of leading
terms such that a two-point correlation can be factored
out, that is
\begin{align}
\langle \tilde{B}(t_{i_1})...\tilde{B}(t_{i_{2n}}) \rangle\approx \langle \mathcal{I}_+  \tilde{B}(t_1)\tilde{B}(t_2)\rangle\langle \mathcal{I}_+  \tilde{B}(t_{k_3})..\tilde{B}(t_{k_{2n}})\rangle\label{newwick}
\end{align}
This approximation can be understood phenomenologically by noting
that two-point
correlation functions $C_{j}(t,t_{1})=\langle \tilde{B}_j(t)\tilde{B}_j
(t_{1})\rangle $ typically decay with the time interval width, e.g., for a
Drude-Lorentzian spectral density, $J(\omega)=2\lambda\gamma\omega/(\omega^2+\gamma^2)$, and at high
temperature, they decay exponentially as $%
e^{-\gamma (t-t_{1})}$. The above approximation (\ref{newwick}) is valid in the limit of large $\gamma$,
however the resulting master equation can capture some non-markovian behavior
because of its time convolution form. 
Note that the relation
(\ref{newwick}) is different from the Bourret approximation for
dichotomic process in which $2n$-point correlations can be exactly
expressed as $\langle \tilde{B}(t_{1})...\tilde{B}(t_{2n})\rangle
=\prod_{k=1}^n\langle\tilde{B}(t_{2k-1})\tilde{B}(t_{2k})\rangle$ \cite{Bourret,Bourret-1}.
Here, the remaining $(2n-2)$ point correlation $\langle\mathcal{I}_+  \tilde{B}(t_{k_3})..\tilde{B}(t_{k_{2n}})\rangle$ in Eq. (\ref{newwick}) still
contains all the permutation of two-point correlation functions given by the expansion (\ref{Wickb}).

We can derive the TC2 equation by utilizing the above
approximations in the general equation of motion for the system density
operator (\ref{cumul}). In appendix A, we obtain the TC2 master equation
\begin{figure}[tp]
\includegraphics[width=8cm,height=6cm]{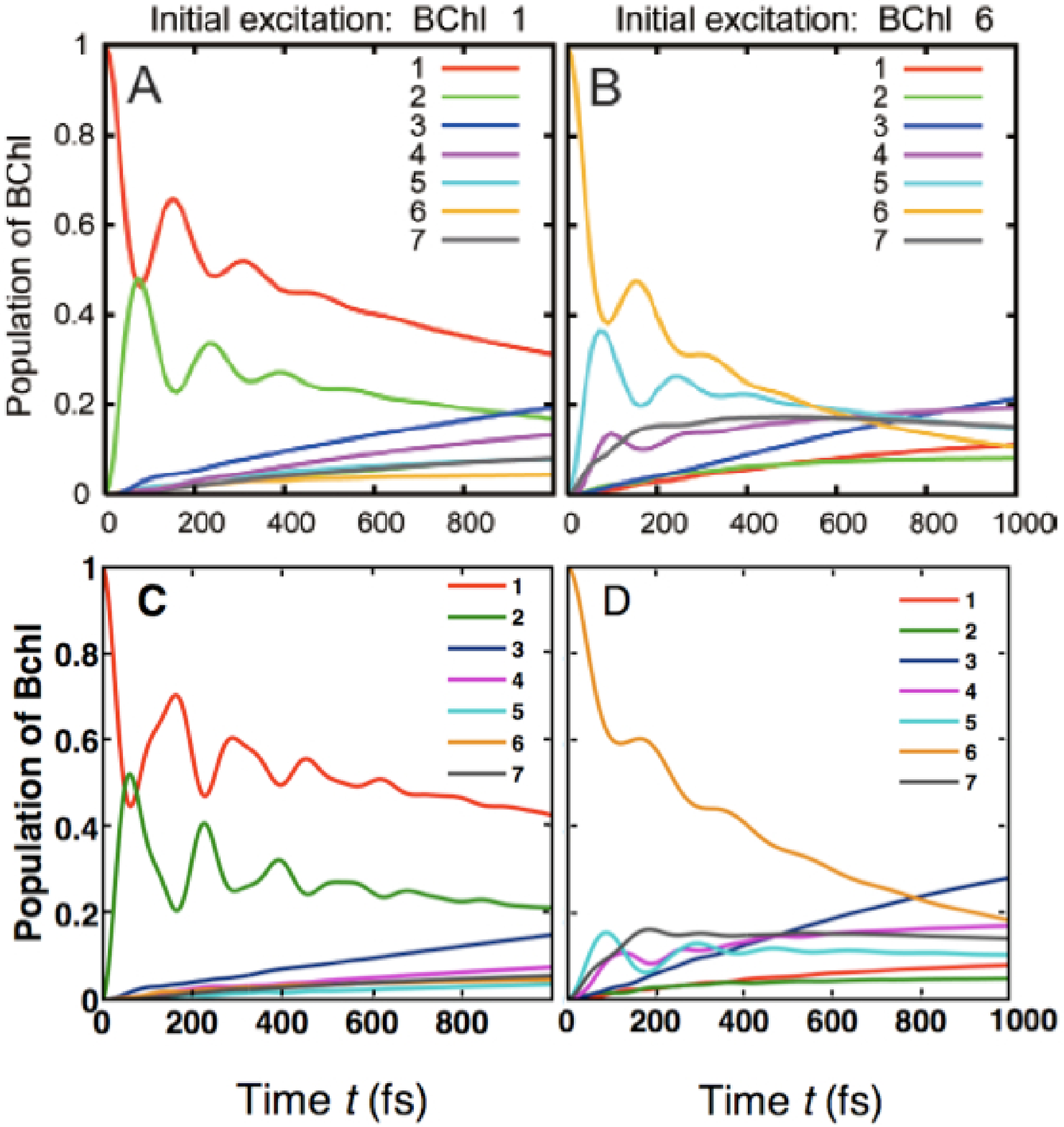}
\caption{Evolution of site populations for all BChls of the FMO
complex at $T=298 K$, $\lambda=35 cm^{-1}$ and $\gamma^{-1}=166 fs$.
The results of simulations using HEOM published in
\cite{AkiPNAS} are shown in top panel (Courtesy of A. Ishizaki). Graphs A and B correspond to
different initial states BChl 1 and BChl 6. The results of the
simulation using TC2 are
presented in graphs C and D are for initial states BChl 1 and BChl
6. This comparison illustrates that the simulation by TC2 yields
oscillatory behavior as those simulations by HEOM approach,
while significantly reducing the computational resources. However
TC2 slightly overestimates amplitudes of oscillations in graph C and
significantly underestimates the decay rates in graph D \cite{Aki-PCCP}. Nevertheless, the measure of energy transfer efficiency that we
employ in this work, as an important yield function to quantify the
performance of light-harvesting systems (see section IV),  is not very sensitive to actual oscillations.} \label{Fig2}
\end{figure}

\begin{eqnarray}
\frac{\partial}{\partial t}\rho(t)=\mathcal{L}_S\rho(t)-\sum_j[S_j,\frac{1}{\hbar^2}\int_0^t C_j(t-t')\times \notag\\
e^{\mathcal{L}_S(t-t')} (S_j\rho(t') ) dt'-h.c.],\label{TCME}
\end{eqnarray}
where $h.c.$ stands for Hermitian conjugate. The derivation of TC2 equation with Born (weak coupling) approximation is well known in the literature \cite{Breuer},
however, in the appendix A, we explicitly derive it under different assumptions away from
weak system-bath limit.
A phase-space representation for the above
equation is also introduced in \cite{Cao} as a generalization of the
Bloch-Redfield equation which is equivalent to
the second tier of hierarchy equations of motion (HEOM) in the high
temperature and Lorentizan spectral density \cite{Ishizaki09-2}. Such generalized Bloch-Redfiled equation is
heuristically concluded from the Gaussian bath assumption which is
inaccurate since it does not account for the pairing operation
in the generalized Wick's expansion (\ref{Wickb}). Here, however, we constructively derive TC2
by starting from the general Liouvillian
equation and introducing the required approximation (\ref{newwick})
without making any weak ambient interactions assumption.   
Here we use TC2 to calculate ETE of a light harvesting complex equipped with
an analysis to estimate the errors introduced
by truncating the expansion (\ref{Wickb}), we find that the master equation can provide reliable estimations in the intermediate
non-perturbative and non-Markovian regimes. An important feature of
the TC2 is that it can be solved efficiently in time using Laplace
transform technique. The numerical complexity of calculating the
corresponding Laplace transform and its inverse does not depend
on the form of the bath spectral density function or the low
temperature limits in contrast to HEOM.

We first use TC2 to simulate the quantum dynamics of the FMO complex (see
Fig. \ref{fmo}) at room temperature against HEOM as a general
benchmark \cite{AkiPNAS}. The FMO Hamiltonian is extracted from
Ref.\cite{Cho}. In both simulations the bath spectral density is
considered to be Drude-Lorentzian with a two time correlation
function of the form $\lambda(2/\beta-i\gamma)e^{-\gamma(t-t_1)}$
where $\gamma$ and $\lambda$ are the cut-off frequency and
reorganization energy, respectively. Instead of solving over two
million coupled differential equations needed for 11 hierarchy
levels, as simulated by Ishizaki and Fleming \cite{AkiPNAS}, here we
only need to solve the time-convolution equation (\ref{TCME}) in the
frequency domain. Fig. \ref{Fig2} shows the oscillatory behavior of
the population of all sites for two different initial conditions and
a comparison with the results of Ref. \cite{AkiPNAS}. This data
suggest that our simulation can capture the essential features of the
FMO dynamics with a significant reduction in computational
resources. It should be noted that our approach leads to slightly
longer oscillations than those in HEOM. This is related to the known side effect of using TC2
yielding a double peak absorption spectra and it appears to be
a generic artifact for any method relying on filtering or truncation
of HEOM \cite{Shi, Patrick2,Aki-PCCP}. As we show below, this issue does not
lead to a major problem in calculating energy
transfer efficiency in which relies on the time average of the
populations is involved. Similar simulations for cryogenic temperature ($T=77 K$)
are presented in appendix B.

\begin{figure}[tp]
\includegraphics[width=8cm,height=5cm]{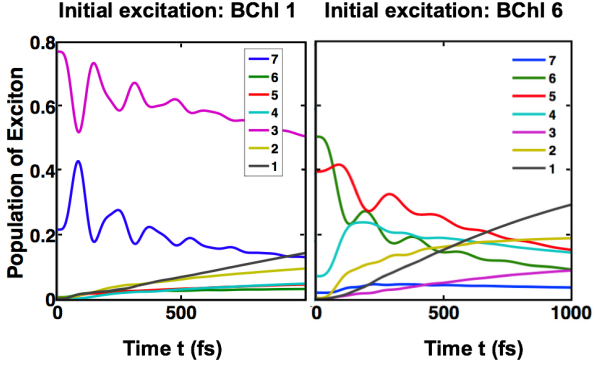}
\caption{Oscillatory dynamics of the FMO complex in the exciton
basis at $T=298 K$, $\lambda=35 cm^{-1}$ and $\gamma^{-1}=166 fs$.
The energy eigenstates are spatially delocalized over various BChls,
thus these oscillations manifest the presence of quantum dynamical
coherence in the FMO complex. We observe that the
exciton oscillations last for a few hundred femtoseconds endorsing
experimental report of relatively long-lived quantum coherence
beating.} \label{shorttimeexciton}
\end{figure}

Our simulation of the dynamics of sites populations in Fig.
\ref{Fig2} illustrated some oscillations that can last for a few
hundred femtoseconds in agreement with the recent experimental results
at room temperature \cite{panit}. However, population oscillations in the site
basis per se cannot confirm the survival of quantum coherence.
To this end, we simulate the dynamics in the exciton
basis corresponding to the eigenvectors of the FMO Hamiltonian. We
present the energy levels populations dynamics over time in Fig.
\ref{shorttimeexciton} indicating hundred of femtoseconds
oscillations. Next, we study
the long-time dynamics of the FMO complex at the picoseconds
time-scales and investigate its the thermal equilibrium properties.
\section{Long-time behavior of excitonic systems}
\begin{figure}[tp]
\includegraphics[width=8cm,height=8cm]{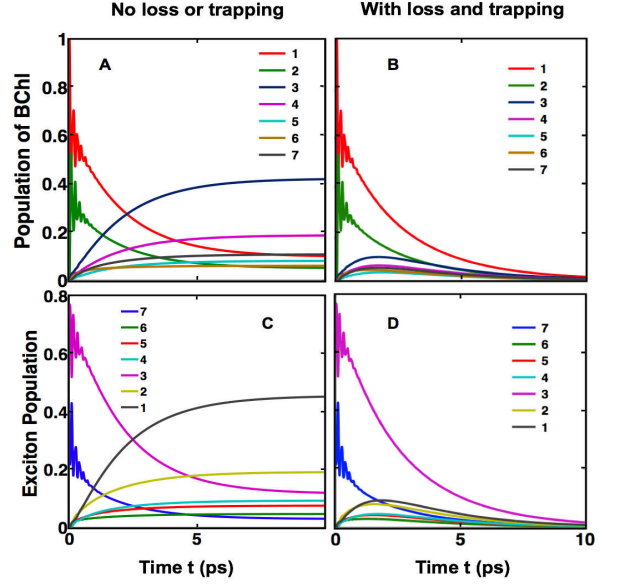}
\caption{Long-time dynamics of the FMO complex for initial
excitation at BChl 1. Left panel presents the simulation without any
lossy mechanisms. In contrast, the right panel illustrates the
results in presence of dissipation and trapping. Top/bottom panel
are associated with dynamics in site and exciton basis. In graphs A
and C, it can be observed that the FMO complex reaches to
equilibrium state within $10 ps$. In graphs (b) and (d) we note that
the exciton is either fully absorbed or lost in the same time-scale
of $10 ps$. Thus, for most part the dynamics, the FMO complex is far
from its equilibrium state.} \label{bchl1quad}
\end{figure}
\begin{figure}[tp]
\includegraphics[width=8cm,height=8cm]{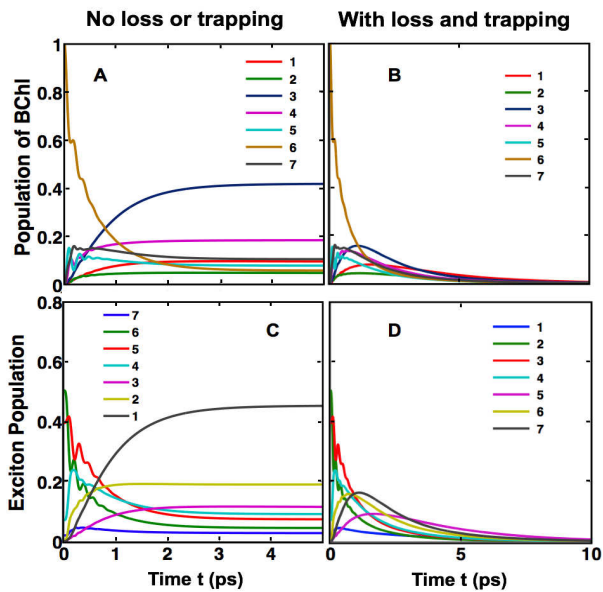}
\caption{Long-time dynamics of the FMO complex for initial
excitation at BChl 6. Left panel presents the simulation without any
lossy mechanisms. The right panel illustrates the results in
presence of dissipation and trapping. Top/bottom panels are
associated with dynamics in site and exciton basis. Plots A and C
illustrate that the FMO equilibrium is achieved within $5 ps$,
faster than the equilibration with initial excitation at BChl 1 as depicted in Fig.
\ref{bchl1quad}. This time difference can be understood by noting
the presence of an energy barrier for excitation transport in the
latter case. In graphs B and D the exciton is mostly absorbed or
lost in less than $7 ps$, again implying non-equilibrium nature of
the transport.} \label{bchl6quad}
\end{figure}
Generally, there are two different competing electron-hole pair
recombination processes that determine the energy transfer
efficiency of light-harvesting systems. The first process happens
within the time-scale of $1$ ns due to dissipation (radiative and non-radiative) to the
environment at each site. This adverse environmental effects
guarantees that the ETE has a value less than one. The second
recombination process is due to successful trapping at one or more
reaction centers that typically occur at the order of picoseconds.
To capture the overall effect of electron-hole pair recombination at
each site, an extra term $\mathcal{L}_{e-h}$ is added to the TC2
(\ref{TNL}).
\begin{eqnarray}
&&\frac{\partial}{\partial t}\rho(t)=\mathcal{L}_S\rho(t)+\mathcal{L}_{e-h}\rho(t) \label{TC2}\\
&&-\sum_j[S_j,\frac{1}{\hbar^2}\int_0^t C_j(t-t')
e^{\mathcal{L}_S(t-t')} (S_j\rho(t') ) dt'-h.c.]\notag
\end{eqnarray}
where $\mathcal{L}_{e-h}=-\sum_j r_{loss}^j\{|j\rangle\langle
j|,.\}- r_{trap}\{|trap\rangle\langle trap|,.\}$. Here
$|trap\rangle$ is state of the site (BChl) connected to the reaction
center. In this paper we assume that the reaction center is
connected to the BChl 3 only: $|trap\rangle=|3\rangle$. The
coefficients $r_{loss}$ and $r_{trap}$ are the recombination and RC
trap rates respectively and $\{,\}$ is the anti-commutator symbol.
Hereon we assume homogenous protein environments that all have
similar correlation functions, $C_j(t-t')=C(t-t')$. Based on the
dynamical equation (\ref{TC2}) we provide a formal definition for
energy transfer efficiency in the next section.

Here, we study the equilibrium state of FMO dynamics using Eq.
(\ref{TC2}). The long-time dynamics of an excitation initially
started at BChl 1 (6) is illustrated in Fig. \ref{bchl1quad}
(\ref{bchl6quad}) in both site and exciton basis. The left panel
shows the dynamics in absence of any electron-hole recombination
processes. The right panel includes the dynamics in presence of both
dissipation and trapping. In the absence of any lossy mechanisms the
system relaxes to a equilibrium state $\rho_1(\infty)$ within $10
ps$ if the exciton starts from BChl1 (see Fig. \ref{bchl1quad} (a)
and (c)) and within $5 ps$ if BChl 6 is the initial state (see Fig.
\ref{bchl6quad} (a) and (c)). The longer equilibration time-scale of
an initial excitation in BChl 1 is a manifestation of an energy
barrier on the exciton path to the trap site BChl 3. Such a barrier
is missing on the exciton transfer from BChl 6 to BChl 3 path
\cite{AkiPNAS}.

It is natural to assume that the FMO complex and vibrational mode of
the scaffold protein are embedded in a thermal bath. Thus the
combined pigment-protein complex should equilibrate to a thermal
Gibbs state. Using TC2, we simulate the infinite time behavior of
FMO and find an equilibrium state very close to its Gibbs state,
denoted by $\rho_G=\exp(-\beta H_S)/Tr(\exp(-\beta H_S))$, for both
initial state BChl 1 and BChl 6: $Tr(|\rho(\infty)-\rho_G|)/2=0.04$.
It should be noted that the true equilibrium state of FMO complex
would be $\rho_{G^*}=Tr_{ph}(\exp(-\beta H_{total})/Tr(\exp(-\beta
H_{total}))$. Therefore only for a very weak system-bath coupling
the FMO steady state becomes $\rho_G$. The TC2 captures this feature by
noting that the distance of the equilibrium state from $\rho_G$
increases from $0.03$ for $\lambda=1 cm^{-1}$ to $0.2$ for
$\lambda=200 cm^{-1}$.

If we include the electron-hole recombination processes due to loss
and trapping, then as we expect the system relaxes to the (zero
excitation) ground state. We observe that the relaxation to the zero
manifold occurs within $10 ps$ that is of the same order of time it
takes for the system to equilibrate if the loss terms are ignored,
see graphs (b) and (d) in Figs. \ref{bchl1quad} and \ref{bchl6quad}.
Thus, for the most part the processes of exciton energy transfer and
trapping occur when the FMO complex is far from its equilibrium
state.

\section{Energy transfer efficiency of light-harvesting systems}

A biologically relevant function for exploring the performance of
light-harvesting complexes is the ETE as defined in Ref.
\cite{mohseni-fmo,Castro08,Ritz}, that is the total exciton
population being successfully trapped.

\begin{eqnarray}
\eta=2r_{trap}\int_0^\infty \langle trap|\rho(t) |trap\rangle dt
\end{eqnarray}
which is simply $2r_{trap}\langle trap|\bar{\rho}(s=0)|trap\rangle$
where $\bar{\rho}(s)$ is the Laplace transform of $\rho(t)$. We
provide a formal derivation of the ETE in appendix C. At each
moment, the overlap of excitonic wave function with the site
connected to the trap, $\langle trap|\rho(t) |trap\rangle$,
quantifies the exciton availability for absorption by the reaction
center. The ETE is indeed a summation over probability of exciton
presence weighted by the trapping rate, therefore a measure of
successful transfer of the exciton captured by the reaction center.

The method developed here allows us to efficiently simulating the
behavior of ETE as a function of various independent system and
environmental degrees of freedom over a wide range of parameters.
Two main parameters characterizing the effect of a Gaussian bath on
an open system are the strength of the system-bath coupling and the
bath internal memory time-scale. The former can be quantified by the
reorganization energy shift $\lambda_j$ which is a quadratic
function of the linear system-bath coupling strength given in
Eq.{\ref{Hamiltonian2}}. The latter is determined by the width of
the phonon modes spectral density (or cut-off frequency) denoted by
$\gamma_j$. Here we assume that all Bchls are interacting with
independent phonon baths with Drude-Lorentzian spectral density
$J_j(\omega)=2\lambda_j\gamma_j\omega/(\omega^2+\gamma_j^2)$.
Although this form of spectral density has been successfully
employed for analyzing experimental results \cite{Read1,Read2,Read3},
theoretical models suggest a
summation of Lorentzian terms with different $\lambda$ and $\gamma$
\cite{McKenzie,McKenzie-1}. We further assume that all baths have equivalent
reorganization energy $\lambda$ and cut-off frequency $\gamma$. This
assumption has been used in several empirical analyses
\cite{Cho,Read1,Read2,Read3,Brixner}. Here we explore the variation of the ETE
versus reorganization energy and bath frequency cutoff. Fig.
\ref{Fig2} demonstrates the optimality of ETE for the FMO protein
complex at room temperature $T=298 K$ and at the
experimentally estimated values of $\lambda=35 cm^{-1}$, $\gamma=50
cm^{-1}$ or $150 cm^{-1}$, assuming trapping and recombination rates as
$r_{trap}^{-1}=1ps$ and $r_{rec}^{-1}=1ns$ respectively consistent
with Refs. \cite{Cho,Ritz,rec,Adolph}. It can be observed that the
memory of the bath can slightly increase ETE in the regimes of weak
system-bath coupling. Moreover, as expected the ETE drops
significantly when the system interacts with a strong and slow bath.
Here, the phenomenon of environment-assisted energy transport that
was first suggested in the context of the Lindblad formalism (weak
coupling and Markovian assumptions) \cite{mohseni-fmo} and
Haken-Strobel-Reineker \cite{Rebentrost08-1,H-S,H-S-1}, can
be observed for all regimes using our approach.
An independent study on the optimality of ETE versus reorganization energy
has been recently reported in Ref. \cite{Wu}.
The role of quantum coherence within the
B800 and B850 rings of LHII in purple bacteria for optimizing energy transfer rates was
also studied earlier by Jang, et. al. using generalized multichromophoric F\"{o}rster
theory Ref.\cite{Jang07}.
We should mention that it takes about 1.8 sec to calculate ETE
using the method presented in this article on a desktop with a 2.4 GHz
processor and a 4 GB RAM memory. Overall, for a quantum system with the Hilbert space size
$d$ the computational cost of simulations using TC2 grows as $\alpha d^{12}$ for some constant $\alpha$.

\begin{figure}[tp]
\includegraphics[width=9cm,height=5cm]{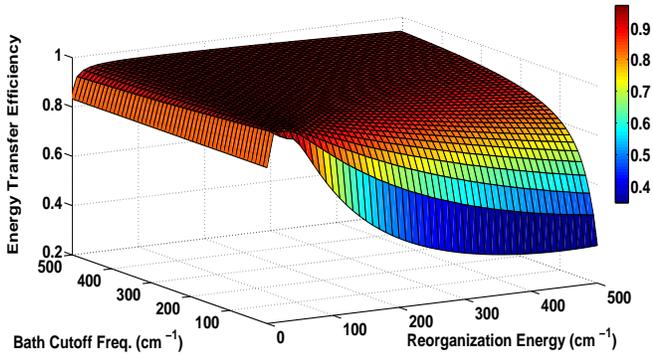}
\caption{Energy transfer efficiency (ETE) of the
Fenna-Matthews-Olson Complex versus reorganization energy $\lambda$
(as a measure of decoherence strength) and bath frequency cutoff
$\gamma$ (as a measure of bath non-Markovianty). The bath frequency
cutoff is plotted starting from $\gamma=5 cm^{-1}$ due large errors
of simulation in highly non-Markov regimes, see Fig. \ref{Fig4}.
The experimentally estimated values of $T=298 K$, $\lambda=35
cm^{-1}$, $\gamma=50 cm^{-1}$ or $150 cm^{-1}$, $r_{trap}^{-1}=1ps$ and
$r_{rec}^{-1}=1ns$ reside at an optimal and robust neighborhood of
ETE. For small reorganization energy (weak system-bath coupling
strength) the non-Markovianity nature of the bath can slightly
increase ETE. However, at larger reorganization energies it will
significantly reduce ETE. Environment-assisted energy transport
\cite{Rebentrost08-2} is clearly observed here.} \label{Fig3}
\end{figure}

\begin{figure}[tp]
\includegraphics[width=9cm,height=6cm]{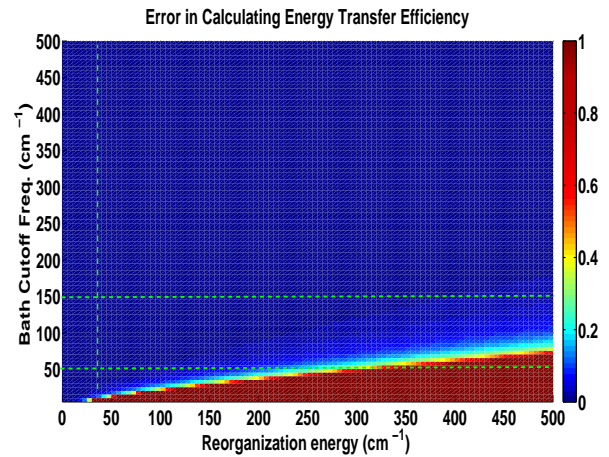}
\caption{The error in calculating the energy transfer efficiency
(ETE) of the FMO complex using TC2: this figure is a complement to Fig. \ref{Fig3} where ETE is
plotted for different values of reorganization energy $\lambda$ and
cut-off frequency $\gamma$. The error estimated by the function
(\ref{totalerror}) increases with a smaller $\gamma$ and a larger
$\lambda$. Overall, the error analysis demonstrates the
reliability of the TC2 for estimating energy transport in excitonic
systems beyond Markovian and perturbative regimes,
specially for the estimated FMO values of $\lambda=35 cm^{-1}$ and
$\gamma=50 cm^{-1}$ or $150 cm^{-1}$ marked by green lines.} \label{Fig4}
\end{figure}

Here we would like to discuss the connection between the
reorganization energy and effective system-bath coupling strength.
The reorganization energy, defined as $\int_0^\infty d\omega J(\omega)/(\pi \omega)$, is usually considered as the single parameter characterizing
the strength of the system-bath coupling. However, in a recent paper \cite{Ritschel}, an insightful observation is made
that this connection is not generally true and one should be careful when interpreting
$\lambda$ as the magnitude of the system interaction with its environment. The reason is a simple fact
that the environmental oscillation modes far from the resonance frequencies of the system do not
contribute to the decoherence dynamics of the system. Ref. \cite{Ritschel} suggests to define an effective
reorganization energy by $\lambda_{eff}=\int_{E_{min}}^{E_{max}} d\omega J(\omega)/(\pi \omega)$
where $[E_{min},E_{max}]$ is the range of the relevant system frequencies. This range is $[0 cm^{-1},550 cm^{-1}] $
for the FMO complex. We have examined this new quantity $\lambda_{eff}$ and considered the ETE as a function
of $\lambda_{eff}$ and $\gamma$. As the simulation results presented in appendix E show, the difference between $\eta(\lambda_{eff},\gamma)$ and $\eta(\lambda,\gamma)$
becomes noticeable in the presence of a weakly coupled Markovian bath.

In a companion paper \cite{comp}, we comprehensively explore the
properties of the FMO complex as a function of all the relevant
environmental and free Hamiltonian parameters. Next, we estimate the
errors introduced into the calculation of ETE using the
time-convolution master equation, and discuss the limitations and
applicability of our approach.

\section{Error estimation of energy transport simulation}

The quantum dynamics as developed above was primarily
motivated to lead to a time non-local master equation (TC2), Eq.
(\ref{TC2}), that may be readily solved in the frequency domain. We
introduced a truncation of the correlation function expansion (\ref{Wickb}) by keeping the
slowly decaying leading terms, and disregarding many fast decaying
higher order bath correlation functions, in order to arrive at a
computationally efficient simulation of quantum dynamics via Laplace
transformation. However, based on the above derivation, the regimes of the applicability of this
method is not clear, since the errors introduced by such a
truncation are not understood quantitatively. Although,
it is qualitatively evident that ignoring such higher order bath
correlation functions will introduce significant errors for very
slow bath and very high reorganization energy. An important issue is
the accuracy of this model in the intermediate regimes. To address
this issue one would ideally attempt to find an exact account of
errors in various regimes of interest. However, this task is
essentially entails calculating the general evolution of the density
operator of the system; that is the exact account of the errors is
computationally as hard as simulating the the exact dynamics of
system in all regimes. Nevertheless, using a combination of
phenomenological and analytical approaches, we provide
error estimation for weak and intermediate system-bath couplings and
bath memory time-scales, thus quantifying reliability and
applicability of our approach in these regimes.

Here, we present an estimate of the ETE calculation inaccuracy associated to our
approximation in Eq. (\ref{newwick}). An upper bound for the
error is
\begin{eqnarray}
\Delta\eta=2r_{trap}|\int_0^\infty \langle
trap|\rho(t)-\rho_{TC2}(t)|trap\rangle dt|\label{errorb}
\end{eqnarray}
where $\rho(t)$ and $\rho_{TC2}(t)$ is the exact density matrix of
the system and $\rho_{TC2}(t)$ is the solution to the TC2.

In order to estimate the above error we need to calculate
$\rho(t)-\rho_{TC2}(t)$. We use the Dyson expansion solutions for
$\tilde{\rho}(t)$ and $\tilde{\rho}_{TC2}(t)$ in the interaction
pictures in the absence of the term $\mathcal{L}_{e-h}$. The effect
of this irreversible term is considered later by introducing a
decaying term $\exp(-r_{trap}t)$ (see appendix D).

\begin{widetext}
\begin{eqnarray}
\tilde{\rho}(t)-\tilde{\rho}_{TC2}(t)=\hspace{4.53in}\notag\\
 \sum_n \int_0^t dt_1...\int_0^{t_{2n-1}} dt_{2n}\sum_{j_1...j_n}\sum_{i_1..i_n}(-1)^{n+k}\Delta_{i_{k+1}..i_{2n}i_{k+1}..i_{k}}\hspace{1.13in}\notag\\
\langle \tilde{B}_{j_{k+1}}(t_{i_{k+1}})..\tilde{B}_{j_{2n}}(t_{i_{2n}})\tilde{B}_{j_1}(t_{i_{1}})..\tilde{B}_{j_k}(t_{i_{k}}) \rangle\times \tilde{S}_{j_1}(t_{i_1})...\tilde{S}_{j_k}(t_{i_k})\rho(0)\tilde{S}_{j_{k+1}}(t_{i_{k+1}})...\tilde{S}_{j_n}(t_{i_n})
\label{difrho1}\hspace{0in}
\end{eqnarray}
\end{widetext}
where $\Delta^{j_1...j_n}_{i_1...i_{2n}}$ is the relative difference between coefficients
of different terms in this expansion
\begin{align}
\Delta^{j_1...j_n}_{i_1...i_n}=1-\frac{\langle \mathcal{I}_+ \tilde{B}_{j_1}(t_{i_1})\tilde{B}_{j_2}(t_{i_2})\rangle\langle \mathcal{I}_+ \tilde{B}_{j_3}(t_{i_3})..\tilde{B}_{j_{2n}}(t_{i_{2n}})\rangle}{\langle \tilde{B}_{j_1}(t_{i_1})...\tilde{B}_{j_{2n}}(t_{i_{2n}}) \rangle}
\end{align}

The above expression for the difference of $\tilde{\rho}(t)$ and
$\tilde{\rho}_{TC2}(t)$ is still exact. The coefficient
$\Delta^{j_1...j_{2n}}_{i_1...i_{2n}}$ quantifies the relative error introduced into a $2n$
bath correlation function by keeping only the slow decaying leading
terms within our approximation. In order to compute the
error in (\ref{errorb}), we first need to estimate the error
contributions from $\Delta^{j_1...j_{2n}}_{i_1...i_{2n}}$.

Here we assume a Drude-Lorentzian bath, however
the analysis can be simply repeated for other types of spectral density functions.
Note that each term ignored in (\ref{Wickb}) decays faster than
the leading  term. In appendix D, we exploit this feature to
arrive at a computable form of  $\Delta^{j_1...j_{2n}}_{i_1...i_{2n}}$  interpolating between
time zero to infinity as $\frac{1}{2n-1}(e^{-\gamma t}+...+e^{-
(2n-2)\gamma t})$. We also use the following inequality \ref{tcorr} to
bound the time-ordered integral of bath correlation functions as
\begin{eqnarray}
|\int_0^tdt_1...\int_0^tdt_n\langle \mathcal{T}_+\tilde{B}
(t_1)...\tilde{B}(t_{2n})\rangle_{ph}| \leq \notag\\ \frac{(2n)!}{2^n n!} |
2\int_0^tdt_1\int_0^{t_1}dt_2\langle \tilde{B}(t_1)\tilde{B}(t_{2})
\rangle_{ph}|^n \label{tcorr}
\end{eqnarray}
where $\frac{(2n)!}{2^n n!}$ is the number of contractions of $2n$ bath operators.
Appendix D shows how to take a further step to estimate the norm
of system operators and arrive at an
expression for an error estimation, defined by Eq. (\ref{errorb}). The estimate for the FMO complex is
\begin{eqnarray}
\Delta\eta=min(1,2r_{trap}\sum_{n}\int_0^\infty dt \frac{\sum_{m=1}^{2n-2}e^{-m\gamma
t}}{2n-1} e^{-r_{trap}t}\times\notag\\
1.75 \frac{\lambda^n}{n!}
(1+4/(\gamma\beta)^2)^{\frac{n}{2}}|t+\frac{1} {\gamma}(e^{-\gamma
t}-1)|^n)\hspace{0.1in}\label{totalerror}
\end{eqnarray}
This estimation is for the high temperature limit $\gamma<\beta^{-1}$. See the
appendix D for the error function in the limit of low temperature
$\gamma>\beta^{-1}$. The behavior of the above error function versus
the reorganization energy $\lambda$ and cut-off frequency $\gamma$
is illustrated in Fig. \ref{Fig4} for a limited region of the ETE
shown in Fig. \ref{Fig3}. This figure shows that the TC2
equation can produce reliable results for intermediate values of the
system-bath coupling and non-Markovian strength. The region in red
denotes the parameter limits in which the TC2 can not produce a
reliable estimate of the ETE based on the error analysis. However as
discussed in appendix D, the function (\ref{totalerror})
overestimates the error and a tighter bound may broaden the
applicability regime of the TC2. The sharp transition from the blue
region (almost zero) error to the red region (almost one) is due to
the convergence/divergence properties of the series in function
(\ref{totalerror}). Note that this error estimation is independent
of the FMO complex properties. A rule of thumb for the applicability
domain of the time nonlocal master equation can be extracted from
Fig. \ref{Fig4}: for a given cut-off frequency $\gamma$, the
system-bath coupling $\lambda$ should satisfy $\lambda\leq c \gamma$
where $c=6$, for $T=298 K$ and $r_{trap}^{-1}=1 ps$. For a detailed
discussion on our error analysis see Appendix D.

\section{Conclusion}
One of the main challenges in understanding the non-equilibrium behavior of photosynthetic complexes and designing artificial harvesting complexes is to efficiently simulate their dynamics when the system-bath coupling is comparable to the system energy scales.
In this work, we have examined the TC2 master equation for simulation of excitonics dynamics
as an efficient tool to compute ETE in complex quantum systems interacting with bosonic environments
in low excitation limits. In particular, we have provided a derivation of the TC2 master
equation without making second-order perturbative assumption. We have applied this
equation to calculate the energy transfer efficiency in
Fenna-Matthews-Olson pigment-protein complex demonstrating optimality and robustness of
energy transfer with respect to variations in bath temporal correlation and reorganization energy. In derivation of the
dynamical equation some approximations have
been made to allow a useful truncation of the $n$-time correlation expansion for
bath time correlations. To account the inaccuracies introduced by
these approximations we have provided an error analysis
estimating the parameter domain for applicability of TC2. 
Specifically, our study here quantify the errors in computing energy transfer efficiency of these systems via TC2. It will be of significant interest to similarly quantify the reliability of alternative schemes for calculations of population transfer such as Modified Redfield theory \cite{modified} or filtering of hierarchy of equations of motions \cite{filtered}.

\begin{acknowledgments}
We thank J. Cao, A. Ishizaki, S. Jang, M. Sarovar, R. Silbey, and K.
B. Whaley for useful discussions. We thank DARPA QuBE program (AS, MM, HR, SL),
NSF (AS, HR), NSERC (MM), and ENI, S.p.A. through the MIT Energy Initiative Program (MM,SL) for funding.
\end{acknowledgments}

\bigskip

\appendix
\section{Derivation of time nonlocal master equation}

The dynamics of a quantum system linearly coupled to a bosonic bath
is generated by the quantum Liouvillian equation
\begin{eqnarray}
\frac{\partial \rho(t)}{\partial t}=-i\hbar\langle[H_{total},\rho(t)]\rangle_{ph}\label{Liouapp}
\end{eqnarray}
with Hamiltonian $H$ given in Eq. (\ref{Hamiltonian}).

The interaction picture solution to Eq. (\ref{Liouapp}) is
\begin{eqnarray}
\tilde{\rho}(t)=\langle \mathcal{T}_+ \exp \Big [\int_0^{t}
\tilde{\mathcal{L}}_{total}(s)ds\Big
]\rangle_{ph}\rho(0),\label{solution}
\end{eqnarray}
where the interaction picture of any operator $O$ is denoted by
$\tilde{O}$. We consider the Dyson expansion of the above
equation \cite{Breuer} (Hereon we drop the subscripts
"$\it{total}$" and "$ph$"):
\begin{widetext}
\begin{align}
\tilde{\rho}(t)= \Big[I+ \int_0^t dt_1\int_0^{t_1} dt_2 \langle \tilde{\mathcal{L}}(t_1)\tilde{\mathcal{L}}(t_2)\rangle 
+\int_0^t dt_1\int_0^{t_1} dt_2 \int_0^{t_2} dt_3 \int_0^{t_3} dt_4 \langle \tilde{\mathcal{L}}(t_1)\tilde{\mathcal{L}}(t_2)\tilde{\mathcal{L}}(t_3)\tilde{\mathcal{L}}(t_4)\rangle+ ... \Big]\rho(0)\label{rhot}
\end{align}
The time derivative of this expansion is:
\begin{align}
\frac{\partial}{\partial t}\tilde{\rho}(t)=\Big[\int_0^t dt_1 \langle \tilde{\mathcal{L}}(t)\tilde{\mathcal{L}}(t_1)\rangle 
+ \int_0^t dt_1\int_0^{t_1} dt_2 \int_0^{t_2} dt_3 \langle \tilde{\mathcal{L}}(t)\tilde{\mathcal{L}}(t_1)\tilde{\mathcal{L}}(t_2)\tilde{\mathcal{L}}(t_3)\rangle+ ... \Big]\rho(0)\label{derive}
\end{align}
\end{widetext}
Note that for $t_1>t_2..>t_n$ the $n$-time correlation superoperator has the following form
\begin{gather}
\langle \tilde{\mathcal{L}}(t_1)...\tilde{\mathcal{L}}(t_n)\rangle\rho(0)=(-i)^n\sum_{j_1..j_n}\sum_{i_1..i_n}(-1)^{n-k}\notag\\
\times\langle \tilde{B}_{j_{k+1}}(t_{i_{k+1}})...\tilde{B}_{j_n}(t_{i_{n}})\tilde{B}_{j_1}(t_{i_{1}})...\tilde{B}_{j_k}(t_{i_k})\rangle \notag\\
\times \tilde{S}_{j_1}(t_{i_1})...\tilde{S}_{j_k}(t_{i_k})\rho(0)\tilde{S}_{j_{k+1}}(t_{i_{k+1}})...\tilde{S}_{j_n}(t_{i_n})\label{nterm}
\end{gather}
with the second summation is over all indices $\{i_1,...,i_n\}\in\{1,..,n\}$ such that the $t_{i_1}>...>t_{i_k}$
and $t_{i_{k+1}}<...<t_{i_n}$, and also with the site indices $\{j_1,...,j_n\}\in\{1,..,N\}$.
Here we assumed that the phonon bath is large enough to
satisfy the Gaussian property; that is the $2m+1$-time correlation
functions vanish and the $2m$-time correlation functions can be
determined by $2$-time correlation functions:
\begin{align}
\langle \tilde{B}(t_{i_1})...\tilde{B}(t_{i_{2m}}) \rangle=\sum_{\substack{all\\pairs}}\prod_{l,k\in i_1..i_{2m}}\langle \mathcal{I}_+\tilde{B}(t_l)\tilde{B}(t_k)\rangle\label{Wicka}
\end{align}
where $\mathcal{I}_+$ is the index ordering operator.
This is a generalized Wick's theorem in the form of Wightman functions \cite{gWick1,gWick2}.
We assume that the bath has a stationary memory
function: $\langle \tilde{B}_j(t_1)\tilde{B}_j(t_2)\rangle=C_j(t_1-t_2)$.

We keep the leading term in the above expansion (\ref{Wicka}):
\begin{align}
\langle \tilde{B}(t_{i_1})...\tilde{B}(t_{i_{2m}}) \rangle\approx \langle \mathcal{I}_+  \tilde{B}(t_1)\tilde{B}(t_2)\rangle\langle \mathcal{I}_+  \tilde{B}(t_{k_3})..\tilde{B}(t_{k_{2m}})\rangle\label{newwick2}
\end{align}
As it will be seen in appendix D, in this approximation we ignore the terms exponentially smaller than
the remaining ones which is valid for relatively large values of $\gamma$.
Note that the relation (\ref{newwick2}) is exact for a cross term due to the assumption of local baths, i.e. for $j\neq j'$
\begin{eqnarray}
\langle \tilde{B}_j(t_{i_1})\tilde{B}_j(t_{i_{2}})\tilde{B}_{j'}(t_{i_3})\tilde{B}_{j'}(t_{i_{4}}) \rangle=\notag\\
\langle \tilde{B}_j(t_{i_1})\tilde{B}_j(t_{i_{2}})\rangle\langle\tilde{B}_{j'}(t_{i_3})\tilde{B}_{j'}(t_{i_{4}}) \rangle
\end{eqnarray}

Now we can factor out $\int_0^t dt_1
\langle\tilde{\mathcal{L}}(t)\tilde{\mathcal{L}}(t_1)\rangle$ from
Eq. (\ref{derive}) and obtain
\begin{eqnarray}
\frac{\partial}{\partial t}\tilde{\rho}_{TC2}(t)=\int_0^t dt_1 \langle \tilde{\mathcal{L}}(t)\tilde{\mathcal{L}}(t_1)\rangle\Big[I +\hspace{0.63in}\notag\\
  \int_0^{t_1} dt_2 \int_0^{t_2} dt_3 \langle \tilde{\mathcal{L}}(t_2)\tilde{\mathcal{L}}(t_3)\rangle+ ... \Big]\rho_{TC2}(0)\notag\\
=\int_0^t dt_1 \langle \tilde{\mathcal{L}}(t)\tilde{\mathcal{L}}(t_1)\rangle \tilde{\rho}_{TC2}(t_1)\hspace{1.03in} \label{TNL}
\end{eqnarray}
The operator $\tilde{\rho}_{TC2}$ represents the state of the system
estimated by TC2.
The above time non-local equation can be simply solved using the
Laplace transform method. Thus, we have:
\begin{align}
\langle \tilde{\mathcal{L}}(t)\tilde{\mathcal{L}}(t_1)\rangle=-\frac{1}{\hbar^2}\sum_j\langle \tilde{B}_j(t)\tilde{B}_j(t_1)\rangle \tilde{S}_j(t)[\tilde{S}_j(t_1),.]-h.c.\notag\\
=-\frac{1}{\hbar^2}\sum_j C_j(t-t_1) \tilde{S}_j(t)[\tilde{S}_j(t_1),.]-h.c.
\end{align}

Now the explicit time convolution form of Eq. (\ref{TNL})
in the Schr\"{o}dinger picture becomes
\begin{align}
\frac{\partial}{\partial t}\rho_{TC2}(t)=\mathcal{L}_S\rho_{TC2}(t)-\sum_j[S_j,\frac{1}{\hbar^2}\times \hspace{1.03in} \notag \\
\int_0^t C_j(t-t')e^{-iH_S(t-t')/\hbar}S_j\rho_{TC2}(t')e^{iH_S(t-t')/\hbar}dt'-h.c.]\label{main}
\end{align}
The above equation describes the influence of the bosonic (phononic)
bath on the dynamics of the system. However, a photosynthesis
complex is typically experiencing three other external dynamical
processes. (a) The interaction with incoming light that generates
the initial exciton in the system. (b) Recombination of the
electron-hole pair at each site (BChl) leading to dissipation of the
exciton to environment. (c) The trapping mechanisms that capture the
excitation energy for charge separation in the reaction center which
eventually converted to biochemical energy. In studying the
excitonic dynamics of a FMO complex the first process is negligible,
since each FMO monomers has a very small absorption cross section
and acts merely as an independent wire to transfer sunlight energy
that is already absorbed by the antenna complex. The two other
irreversible processes are modeled by adding two corresponding lossy
terms to the RHS of Eq. (\ref{main}) as:
\begin{widetext}
\begin{align}
\frac{\partial}{\partial t}\rho_{TC2}(t)=\mathcal{L}_S\rho_{TC2}(t)+\mathcal{L}_{e-h}\rho_{TC2}(t)
-\sum_j[S_j,\frac{1}{\hbar^2}\int_0^t C_j(t-t')e^{-iH_S(t-t')/\hbar}S_j\rho_{TC2}(t')e^{iH_S(t-t')/\hbar}dt'-h.c.]\label{lossymain}
\end{align}
\end{widetext}
where $\mathcal{L}_{e-h}=-\sum_j r_{rec}^j\{|j\rangle\langle
j|,.\}-r_{trap}\{|trap\rangle\langle trap|,.\}$ with $r_{rec}$
($r_{trap}$) being the recombination (RC trap) rate and
$|trap\rangle$ represents the state of the BChl connected to the RC.
In this article we use Eq. (\ref{lossymain}) as the main dynamical
equation describing the excitonic energy transfer process in the FMO
complex.

 It should be noted that the well-known derivation of TC2 is based on the
assumption that system-bath coupling is so weak such that the
third and higher order terms of the expansion (\ref{rhot}) and (\ref{derive}) can be
ignored. Basically, one can solve (\ref{rhot}) to find $\rho(0)$ as a function of $\tilde{\rho}(t_1)$
and substitute it in Eq. (\ref{derive}), all up to the second orders of $\mathcal{L}$, and obtain TC2.
In this paper we avoid the assumption of the weak
system-bath coupling and we keep all powers of the reorganization
energy in (\ref{derive}). Instead, here, we introduce the
approximation (\ref{newwick2}) to arrive at Eq.(\ref{main}).
Notice that in our approach the expanded solution of the density matrix
has not been truncated over a finite power series of a
"small" physical parameter (e.g. site energies, dipole-dipole interactions,
reorganization energy, or system-bath Hamiltonain parameters in any
fixed or rotating reference frame). 

Finally we would like to clarify that the approximation
(\ref{newwick2}) is different from the Bourret approximation for
which the Eq. (\ref{newwick2}) is applied recursively that is the
$(2n-2)$ correlation function on RHS is also approximated \cite{Bourret,Bourret-1}. Note that
in contrast we consider a Gaussian bath and apply the approximation
(\ref{newwick2}) separately to each term of the expansion
(\ref{derive}) and thus treating the $(2n-2)$-point correlation
functions on RHS of Eq. (\ref{newwick2}) exactly. That is in our
approximation we keep one out of the overall 2n terms in the Wick's
expansion for the final time step. On the other hand, within the
Bourret approximation, one attempts to capture the dynamics by
keeping only one of the $\frac{(2n)!}{2^n n!}(\approx (2n/e)^n$ for large $n)$ terms for two-point correlation
functions. Nevertheless, the Bourret approximation become exact when
the noise can be regarded as a dichotomic random process leading to
$\langle\tilde{B}(t_{1})...\tilde{B}(t_{2n})\rangle
_{ph}=\prod_{k=1}^n\langle\tilde{B}(t_{2k-1})\tilde{B}(t_{2k})\rangle
_{ph}$. It remains an open problem to express the dynamics of the
bath as a physically motivated random process leading to the relation (\ref{newwick2}).
For a detailed error analysis of our approximation for computing
energy transfer efficiency see appendix D.

\section{Excitonic Dynamics at Cryogenic Temperature}
\begin{figure}[tp]
\includegraphics[width=8cm,height=6cm]{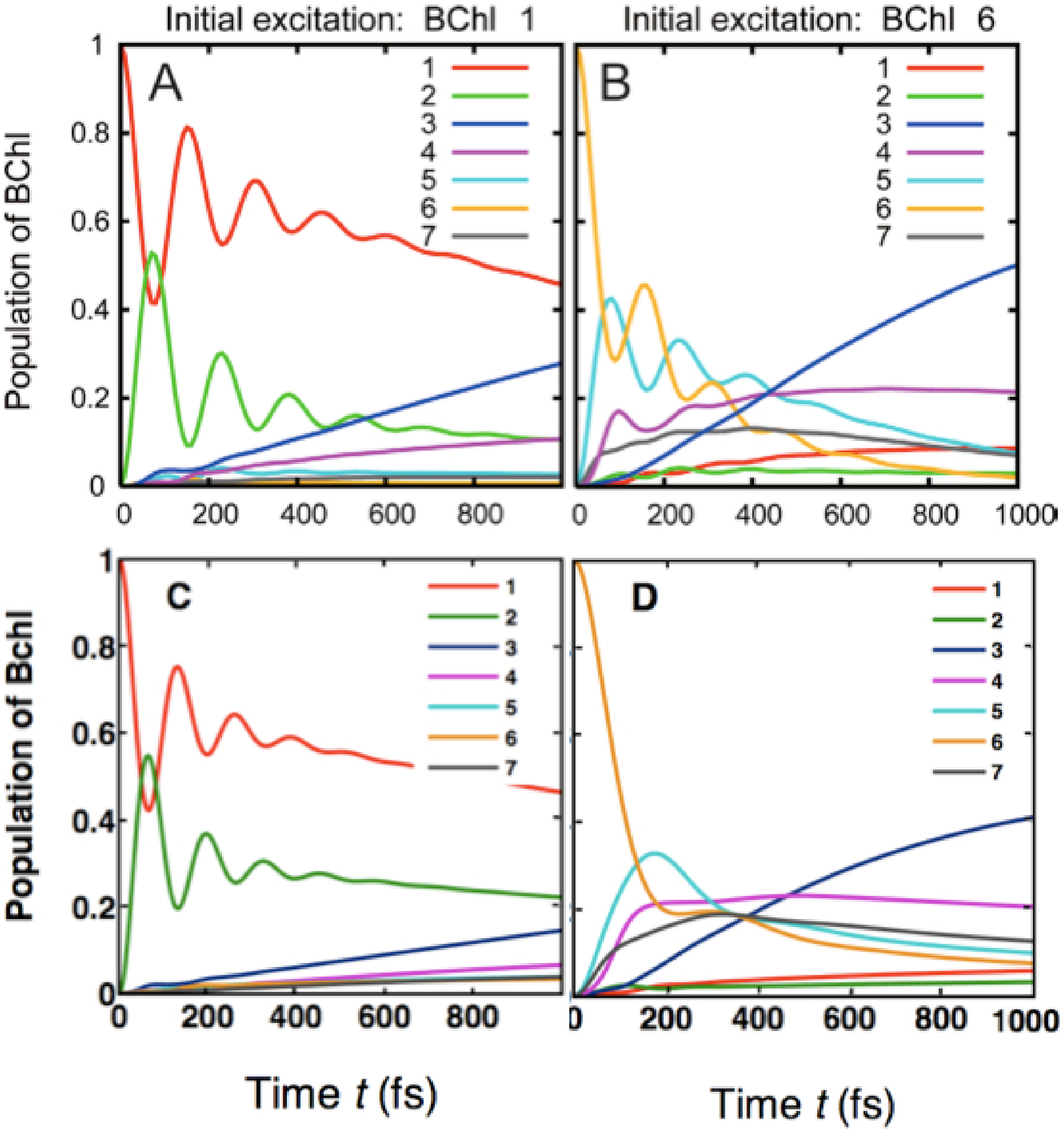}
\caption{Evolution of site populations for all BChls of the FMO
complex at $T=77 K$, $\lambda=35 cm^{-1}$ and $\gamma^{-1}=50 fs$.
The results of simulations using HEOM published in
\cite{AkiPNAS} are shown in top panel (Courtesy of A. Ishizaki). Graphs A and B correspond to
different initial states BChl 1 and BChl 6. The results of the
simulation using TC2 are
presented in graphs C and D are for initial states BChl 1 and BChl
6. TC2 predicts faster damping with initialization in BChl 1 and almost
no oscillation for BChl 6 as the initial state.} \label{Fig77}
\end{figure}

We simulated the excitonic dynamics of FMO at $T=77 K$ using TC2. Notice that here we consider $\gamma^{-1}=50 fs$
and $\lambda=35 cm^{-1}$ to make a comparison with simulations in Ref. \cite{AkiPNAS}.
At this temperature the bath two-time correlation function can be well approximated
by $C(t-t_1)=\lambda(\frac{2}{\beta}(1-\frac{2\gamma^2}{\nu^2-\gamma^2})-i\gamma)e^{-\gamma(t-t_1)}+\frac{4\lambda\gamma}{\beta(\nu^2-\gamma^2)}\delta(t)$,
where $\nu=\frac{2\pi}{\beta\hbar}$ \cite{AkiPNAS}. Fig. (\ref{Fig77}) illustrates the
simulation of BChls population for both initial state BChl 1 and BChl 6.
Notice that in Fig. (\ref{Fig77}), $\gamma$ is almost three times larger than in the
Fig. (\ref{Fig2}), approaching Markovian regime. In comparison to
HEOM results, TC2 underestimates the oscillations for initial
excitation at BChl 1 and predicates negligible oscillations for initial excitation at BChl 6.

\section{Derivation of Energy Transfer Efficiency Function}

Generally, the sunlight energy captured by the antenna complexes has
to be transferred to one or more reaction centers for storage.
However, there is always some finite chance of radiative or
nonradiative electron-hole recombination at each BChl sites leading
to dissipation of energy to environment as florescence or quenching.
The amount of the initial exciton that eventually arrives at RC
determines the efficiency of the energy transfer process.

Equation (\ref{lossymain}) captures the dynamical evolution of the
system in the single-exciton manifold. However, for a complete
picture we consider the dynamical equation for the full quantum
state $\rho_{full}$ over an extended Hilbert space. This
$(N+2)$-dimensional extended Hilbert space is constructed from $N$
energy levels in single excitation manifold, the single state,
$|0\rangle$, in zero excitation manifold, and finally a single state
$|RC\rangle$ representing the reaction center. Modeling the presence
of the reaction center by just an additional state is allowed since
only exciton population and not coherence is transferred from the
system to the reaction center.
\begin{widetext}
\begin{align}
\frac{\partial}{\partial t}\rho_{full}(t)=-i\mathcal{L}_S\rho_{full}(t)+\mathcal{L}_{e-h}^{full}\rho_{full}(t)
-\sum_j[S_j,\frac{1}{\hbar^2}\int_0^t C_j(t-t')e^{-iH_S(t-t')/\hbar}S_j\rho_{full}(t')e^{iH_S(t-t')/\hbar}dt'-h.c.]\label{loss1}
\end{align}
\end{widetext}
where
\begin{align}
\mathcal{L}_{e-h}^{full}\rho=\sum_jr_{rec}^j\Big [-\{|j\rangle\langle j|,\rho\}+2 |0\rangle\langle j|\rho|j\rangle\langle 0|\Big]+\notag\\
r_{trap}\Big [-\{|trap\rangle\langle trap|,\rho\}+2 |RC\rangle\langle trap|\rho|trap\rangle\langle RC|\Big]\label{loss2}
\end{align}
The total amount of the initial exciton population finally trapped in the reaction center is $\langle RC|\rho_{full}(\infty)|RC\rangle$.
 This can be equivalently evaluated by using Eqs. (\ref {loss1}) and (\ref {loss2}) as
\begin{equation}
\frac{\partial}{\partial t}\langle RC|\rho_{full}(t)|RC\rangle=2r_{trap} \langle trap|\rho_{full}(t)|trap\rangle
\end{equation}
therefore
\begin{align}
\eta=\langle RC|\rho_{full}(\infty)|RC\rangle=2r_{trap}\int_{0}^{\infty}dt \langle trap|\rho_{full}(t)|trap\rangle \notag\\
=2r_{trap}\int_{0}^{\infty}dt \langle trap|\rho_{TC2}(t)|trap\rangle\label{eta}
\end{align}
Note that the RHS of the above equation is equivalent to $2r_{trap}\langle trap|\bar{\rho}(s=0)|trap\rangle$, where $\bar{\rho}(s)$ is the Laplace transform of $\rho(t)$.
The convolutional form of the Eq. (\ref{lossymain}) allows simple calculation of the $\bar{\rho}(s)$, which yields a closed form for the
energy transfer efficiency function.

The Laplace transformed form of the main equation (\ref{main}) is
\begin{eqnarray}
s\bar{\rho}(s)-\rho(0)&=&\mathcal{L}_S\bar{\rho}(s)+\mathcal{L}_{e-h}\bar{\rho}(s) \notag\\
&-&\sum_j[S_j,\bar{K}^j(s)-\bar{K}^{j\dagger}(s)]
\end{eqnarray}
where
\begin{eqnarray}
\bar{K}^j(s)=\frac{1}{\hbar}\bar{C}(s+i\mathcal{L}_S)(S_j\bar{\rho}(s)).\notag\label{Lap}
\end{eqnarray}
We can find the efficiency function $\eta$ by solving this equation for $\bar{\rho}(s=0)$.

\section{Error analysis for simulation of transport efficiency}

In the previous sections, we showed that the ETE of an excitonic system can be
calculated efficiently by combining the TC2 and Laplace
transform technique. However, this computational simplicity inherently induces errors associated
with ignoring certain higher order bath correlation functions.

In this section we estimate the error in calculating the efficiency function $\eta$ (\ref{eta}).
An upper bound for the error is defined as:

\begin{eqnarray}
\Delta\eta=2r_{trap}|\int_0^\infty \langle trap|\rho(t)-\rho_{TC2}(t)|trap\rangle dt| \notag \\
\leq 2r_{trap}\int_0^\infty |\langle trap|\rho(t)-\rho_{TC2}(t)|trap\rangle|dt\label{errorwithourtrap}
\end{eqnarray}

The exact dynamical equation for the system in the presence of recombination and trapping is obtained from
Eq. (\ref{Liou}) by adding the term $\mathcal{L}_{e-h}\rho(t)$
\begin{eqnarray}
\frac{\partial \rho(t)}{\partial t}=\mathcal{L}_{e-h}\rho(t)-i\hbar\langle[H_{total},\rho(t)]\rangle\label{Liou2}
\end{eqnarray}
An interaction picture solution to Eq. (\ref{Liouapp}) is given in (\ref{rhot}), but the term $\mathcal{L}_{e-h}\rho(t)$
will make it difficult to find a compact solution to Eq. (\ref{Liou2}). This complexity arises from the non-invertibility
of the loss propagator operator $\exp(\mathcal{L}_{e-h}t)$. Instead, we consider $\rho(t)$ and $\rho_{TC2}(t)$
to be solutions to Eqs. (\ref{Liouapp}) and (\ref{main}), respectively, which do not include the effect of loss due to recombination and trapping in the reaction center.
We incorporate these effects heuristically by considering that both $\rho(t)$ and $\rho_{TC2}(t)$ decay as $\exp(-r_{trap}t)$.
The new function to evaluate is
\begin{align}
\Delta\eta=2r_{trap}\int_0^\infty dt e^{-r_{trap}t}|\langle trap|\rho(t)-\rho_{TC2}(t)|trap\rangle|
\end{align}

The TC2 (\ref{main}) can be solved
for $\tilde{\rho}_{TC2}(t)$ yielding
\begin{widetext}
\begin{eqnarray}
\tilde{\rho}_{TC2}(t)&=& \Big[I+ \int_0^{t} dt_1\int_0^{t_1} dt_2 \langle \tilde{\mathcal{L}}(t_1)\tilde{\mathcal{L}}(t_2)\rangle \notag \\
&+& \int_0^t dt_1\int_0^{t_1} dt_2 \int_0^{t_2} dt_3 \int_0^{t_3}
dt_4 \langle\tilde{\mathcal{L}}(t_1)\tilde{\mathcal{L}}(t_2)\rangle\langle
\tilde{\mathcal{L}}(t_3)\tilde{\mathcal{L}}(t_4)\rangle+
... \Big]\rho(0)\label{rhotnl2}
\end{eqnarray}
and the difference between the exact and approximate solutions is:

\begin{align}
\tilde{\rho}(t)-\tilde{\rho}_{TC2}(t)=\hspace{4.53in}\notag\\
 \sum_n \int_0^t dt_1...\int_0^{t_{2n-1}} dt_{2n}\sum_{j_1...j_n}\sum_{i_1..i_n}(-1)^{n+k}\Delta^{j_{k+1}..j_{2n}j_{k+1}..j_{k}}_{i_{k+1}..i_{2n}i_{k+1}..i_{k}}\hspace{1.13in}\notag\\
\langle \tilde{B}_{j_{k+1}}(t_{i_{k+1}})..\tilde{B}_{j_{2n}}(t_{i_{2n}})\tilde{B}_{j_1}(t_{i_{1}})..\tilde{B}_{j_k}(t_{i_{k}}) \rangle\times \tilde{S}_{j_1}(t_{i_1})...\tilde{S}_{j_k}(t_{i_k})\rho(0)\tilde{S}_{j_{k+1}}(t_{i_{k+1}})...\tilde{S}_{j_n}(t_{i_n})\label{difrho}\hspace{0in}
\end{align}
where
\begin{eqnarray}
\Delta^{j_1...j_n}_{i_1...i_n}=1-\frac{\langle \mathcal{I}_+ \tilde{B}_{j_1}(t_{i_1})\tilde{B}_{j_2}(t_{i_2})\rangle\langle \mathcal{I}_+ \tilde{B}_{j_3}(t_{i_3})..\tilde{B}_{j_{2n}}(t_{i_{2n}})\rangle}{\langle \tilde{B}_{j_1}(t_{i_1})...\tilde{B}_{j_{2n}}(t_{i_{2n}}) \rangle}.
\end{eqnarray}
\end{widetext}

Here we continue the analysis for a Drude-Lorentzian bath at high-temperature $\beta^{-1}$, with cutoff frequency $\gamma$ and reorganization energy $\lambda$, which
is the case for FMO complex: $C(t-t_1)=\langle \tilde{B}(t)\tilde{B}(t_1)\rangle=\lambda(2/\beta\pm i\gamma)e^{-\gamma(t-t_1)}$ ($\pm$ sign is determined by the order of $t$ and $t_1$). The analysis can be repeated for other types
of spectral density functions.

Each term ignored in Eq. (\ref{nterm}) decays faster than the leading term
\begin{widetext}
\begin{eqnarray}
|\langle\mathcal{I}_+ \tilde{B}(t_1)\tilde{B}(t_2)\rangle\langle \mathcal{I}_+ \tilde{B}(t_3)...\tilde{B}(t_{2n})\rangle|>|\langle\mathcal{I}_+ \tilde{B}(t_1)\tilde{B}(t_k)\rangle\langle \mathcal{I}_+ \tilde{B}(t_2)...\tilde{B}(t_{k-1})\tilde{B}(t_{k+1})...\tilde{B}(t_{2n})\rangle|
\end{eqnarray}

This behavior is evident in time ordered correlation terms 

\begin{eqnarray}
\langle \tilde{B}(t)...\tilde{B}(t_3) \rangle&=&C(t-t_1)C(t_2-t_3)+C(t-t_2)C(t_1-t_3)+C(t-t_3)C(t_1-t_2) \notag\\
&=&\lambda^2(2/\beta-i\gamma)^2e^{-\gamma(t-t_3)}[e^{\gamma(t_1-t_2)}+2e^{-\gamma(t_1-t_2)}]\label{five}
\end{eqnarray}
\end{widetext}
where we approximate $e^{\gamma(t_1-t_2)}+2e^{-\gamma(t_1-t_2)}$ with $e^{\gamma(t_1-t_2)}$.
A similar calculation can be done for higher order terms.
This approximation introduces larger errors for smaller cut-off frequencies $\gamma$.
Thus, stronger non-Markovian characteristic of the bath results in higher inaccuracy.
For short times $t$ we can make an estimation by assuming that all the terms in the expansion (\ref{five}) have the same magnitude:
\begin{align}
&|C(t-t_1)C(t_2-t_3)+C(t-t_2)C(t_1-t_3)\notag \\
&+C(t-t_3)C(t_1-t_2)|\sim 3 |C(t-t_1)C(t_2-t_3)|
\end{align}
or for the $2n$'th terms
\begin{align}
&|C(t-t_1)C(t_2,t_3,...)+C(t-t_2)C(t_1,t_3,...)\notag \\
&+C(t-t_3)C(t_1,t_2,...)|\sim (2n-1) |C(t-t_1)C(t_2,t_3,...)|
\end{align}
This implies that the coefficient $\lambda^{2n}$ is renormalized as $\lambda^{2n}/(2n-1)$. This approximation becomes exact for long times.
Therefore the coefficient error $\Delta$ goes from $\frac{2n-2}{2n-1}$ at short-times to the value of zero at long-times.
Here we construct functions $\Delta$ interpolating between short and long times.
To this end, we make the following assumption: 

\begin{align}
&|C(t-t_{k+1})C(t_1,...,t_{k},t_{k+2},...,t_{2n-1})|_{ave}\notag \\
&<e^{-\gamma t}|C(t-t_k)C(t_1,...,t_{k-1},t_{k+1},...,t_{2n-1})|_{ave},\label{condition1}
\end{align}
, for time $t$, and on average for different values of $t_1,...,t_{2n-1}$, that yields
\begin{align}
|C(t-t_k)C(t_1,...,t_{k-1},t_{k+1},...,t_{2n-1})|_{ave}\notag \\
<e^{-(k-1)\gamma t}|C(t-t_1)C(t_2,t_3,...)|_{ave}.
\end{align}
Note that this assumption is additional to the Gaussian properties of the bath, and it holds if higher order bath correlation
terms in the expansion (\ref{Wicka}) decay as an exponential function of the bath cut-off frequency $\gamma$.
Using this relation, we find a computable form for the error function $\Delta$ that is $\frac{1}{2n-1}(e^{-\gamma t}+...+e^{-(2n-2)\gamma t})$
interpolating between $\frac{2n-2}{2n-1}$ at time zero to zero at time infinity. We substitute this expression into (\ref{difrho}) and find an estimate for the efficiency error
\begin{widetext}
\begin{eqnarray}
\Delta\eta&=&2r_{trap}\sum_{n>1}\int_0^\infty dt \frac{e^{-\gamma t}+...+e^{-(2n-2)\gamma t}}{2n-1} e^{-r_{trap}t}\times\notag \\
&& \frac{1}{(2n)!}|\int_0^tdt_1\int_0^t...\int_0^tdt_n\mathcal{T}_+\sum_{i_1..i_{2n}}\langle \tilde{B}(t_{i_{k+1}})...\tilde{B}(t_{i_{2n}})\tilde{B}(t_{i_1})...\tilde{B}(t_{i_k})\rangle| \notag \\
&&\sum_{j_1...j_{2n}}| \langle trap|\exp(-iH_{S}t)\mathcal{T}_+\tilde{S}_{j_1}(t_{i_1})...\tilde{S}_{j_k}(t_{i_k})\rho(0)\tilde{S}_{j_{k+1}}(t_{i_{k+1}})...\tilde{S}_{j_{2n}}(t_{i_{2n}})\exp(iH_{S}t)|trap\rangle|
\label{error-bound}
\end{eqnarray}
\end{widetext}
Here we used the identity: $\int_0^tdt_1...\int_0^{t_{n-1}}dt_n
\tilde{\mathcal{L}}(t_1)...\tilde{\mathcal{L}}(t_n)=\frac{1}{n!}\int_0^tdt_1...\int_0^tdt_n\mathcal{T}_+\tilde{\mathcal{L}}(t_1)...\tilde{\mathcal{L}}(t_n)$.
The time ordering operator, $\mathcal{T}_+$, is not very suitable for
our purpose here. This operator is ordering the superoperators
$\tilde{\mathcal{L}}(t)$ and not the bath or system operators
$\tilde{B}(t)$ or $\tilde{S}(t)$. Instead the time ordering of
$\tilde{\mathcal{L}}(t)$s imposes a more complicated ordering of
$\tilde{B}(t)$ or $\tilde{S}(t)$s as described in the expansion
(\ref{nterm}).

Next we consider an estimate for
\begin{widetext}
\begin{eqnarray}
Z^{j_1...j_n}_{i_1..i_n}=\sum_{j_1...j_n}| \langle trap|\exp(-iH_{S}t)\mathcal{T}_+\tilde{S}_{j_1}(t_{i_1})...\tilde{S}_{j_k}(t_{i_k})\rho(0)\tilde{S}_{j_{k+1}}(t_{i_{k+1}})...\tilde{S}_{j_{2n}}(t_{i_{2n}})\exp(iH_{S}t)|trap\rangle|. \label{Zfunction}
\end{eqnarray}
\end{widetext}
Each operator $\tilde{S}(t_k)$ in the interaction picture equals $\exp (-iH_{S}t_k)S\exp (iH_{S}t_k)$.
Estimating this bound is not tractable in general, instead here we make an intuitive estimate. The coefficient $Z$ is a product of the terms $L(j,j')=|\langle j|\exp (-iH_{S}t_k)|j'\rangle|=\sqrt{\sum_\alpha|\langle j|\psi_\alpha\rangle|^2|\langle j'|\psi_\alpha\rangle|^2}$ where $|\psi_\alpha\rangle$s are exciton basis. This expression represents how delocalized a site state can become over time. For a system with small terms $L(j,j'\neq j)$ we can ignore the cross terms in (\ref{Zfunction}). This is the case for FMO with average $L(j,j'\neq j)=0.05$ \cite{Cho}. From the average value of the terms $L(j,j)=L$ we estimate $Z\sim L^{2n+2}$. We consider $L=0.5$ for FMO denoting that the wavefunction on average becomes delocalized over two sites. In addition note that there are $2^{2n}$ different indices $\{i_1..i_n\}$ and $N$ number of sites.

By applying the Gaussian property (\ref{Wicka}) one can find \cite{Ng2}
\begin{eqnarray}
|\int_0^tdt_1\int_0^t...\int_0^tdt_n\langle \mathcal{T}_+\tilde{B}(t_1)...\tilde{B}(t_{2n})\rangle| \notag \\
 \leq \frac{(2n)!}{2^n n!} |2\int_0^tdt_1\int_0^{t_1}dt_2\langle \tilde{B}(t_1)\tilde{B}(t_{2})\rangle|^n \notag \\
= \frac{(2n)!}{n!} |\lambda\sqrt{4/(\gamma\beta)^2+1}(t+\frac{1}{\gamma}(e^{-\gamma t}-1))|^n.\label{tcorr2}
\end{eqnarray}
where $\frac{(2n)!}{2^n n!}$ is the number of contractions of $2n$ operators $\tilde{B}(t_{1})$ to $\tilde{B}(t_{2n})$.

We now substitutes $Z$ and (\ref{tcorr2}) into Eq. (\ref{error-bound}) to find the final expression for the error estimate
\begin{eqnarray}
\Delta\eta=2r_{trap}\sum_{n>1}\int_0^\infty dt \frac{\sum_{m=1}^{2n-2}e^{-m\gamma
t}}{2n-1} e^{-r_{trap}t}\times \notag \\
N4^nL^{2n+2} \frac{\lambda^n}{n!}
(1+4/(\gamma\beta)^2)^{\frac{n}{2}}|t+\frac{1} {\gamma}(e^{-\gamma
t}-1)|^n \label{errorestimation}
\end{eqnarray}
\begin{figure*}[tp]
\includegraphics[width=15cm,height=5cm]{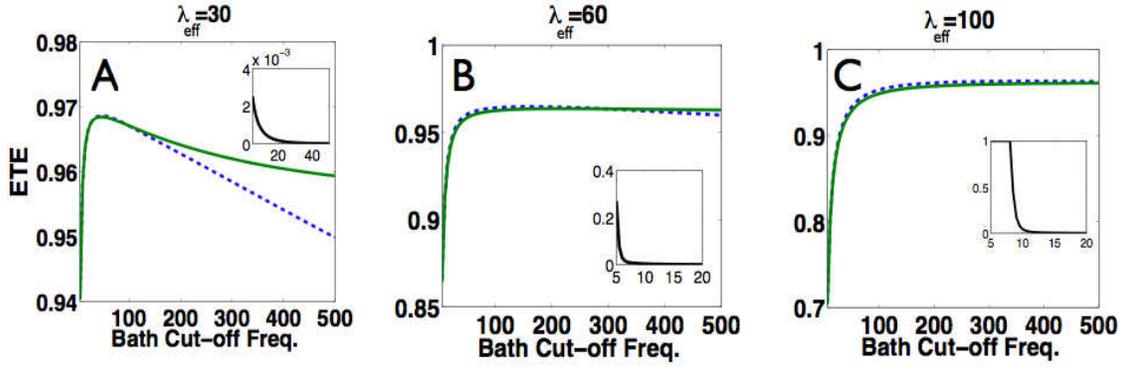}
\caption{The ETE as function of the effective reorganization energy. The sold (dashed) line shows ETE for a fixed value of $\lambda_{eff}$ ($\lambda$) for three different values of $30$, $60$ and $100$ $cm^{-1}$.
The difference between $\eta(\lambda_{eff},\gamma)$ and $\eta(\lambda,\gamma)$ becomes noticeable in the regime of large $\gamma$ and small $\lambda$. The insets show the error estimation versus bath cutoff frequency for the fixed effective reorganization energy.} \label{effectivelambda}
\end{figure*}
\section{Effective Reorganization Energy}
In a recent study \cite{Ritschel}, Ritschel et al. introduce an effective reorganization energy as a measure for the
system-bath coupling strength following the fact that the bath modes which are off-resonance with the system frequencies have negligible
effect on the dynamics of the system therefore the definition of the reorganization energy $\lambda_{eff}=\int_{0}^{\infty} d\omega J(\omega)/(\pi\omega)$ should be modified as
$\lambda_{eff}=\int_{E_{min}}^{E_{max}} d\omega J(\omega)/(\pi\omega)$, where $E_{max}$ and $E_{min}$ are the maximum and minimum system frequencies.
For the Lorentzian spectral density, $J(\omega)=2\lambda\gamma\omega/(\omega^2+\gamma^2)$, and $[E_{min},E_{max}]=[0cm^{-1},550cm^{-1}]$
we find $\lambda_{eff}=(2\lambda/\pi) tan^{-1}(550/\gamma)$. It can be seen from the notion of an effective reorganization energy that the bath modes and therefore a cut-off frequency $\gamma$
beyond $550cm^{-1}$ are less relevant for the FMO excitonic dynamics.
We have plotted the ETE and the estimated error for fixed effective reorganization energies in Fig. (\ref{effectivelambda}).
A one percent difference between $\eta(\lambda_{eff},\gamma)$ and $\eta(\lambda,\gamma)$ plots is observed in the regime of large $\gamma$ and small $\lambda$, as was expected.

The energy transfer efficiency by definition (\ref{eta}) has a
maximum value of one. Therefore, any erroneous estimation of the
efficiency should be less than one. However, in the above
analysis the error is overestimated due to summing the magnitude
of all the terms. That might cause errors larger than one in some regimes which we
reexpress them with a maximum error value of one. A similar approach is taken to
estimate the noise threshold for fault-tolerant quantum computation
in the presence of Gaussian noise \cite{Ng2,Ng}, where the threshold is found
based on bath two-time correlation functions.

It should be noted that the above analysis has been done for high temperature limit characterized by
$\gamma<\beta^{-1}$. In the landscape study presented in this paper, we also consider $\gamma$ values greater than $\beta^{-1}$
for which some correction terms need to be added to the bath correlation function.
For $T=298^{\circ}K$ and $\gamma<500$ $cm^{-1}$, it is enough to consider the following expression as the bath correlation function \cite{AkiPNAS,AkiLowT}
\begin{align}
\langle\tilde{B}(t)\tilde{B}(t')\rangle_{ph}=\lambda(\frac{2}{\beta}-\frac{4\gamma^2/\beta}{(2\pi/\beta)^2-\gamma^2}-i\gamma)e^{-\gamma (t-t')}\notag\\
+\frac{4\gamma^2/\beta}{(2\pi/\beta)^2-\gamma^2}\delta(t-t')
\end{align}
Including this new correlation function, the error estimation can be
simply obtained by replacing the coefficient $(1+4/(\gamma\beta)^2)$
with
$(1+(\frac{2}{\gamma\beta}-\frac{4\gamma\beta}{(2\pi)^2-(\beta\gamma)^2})^2)$
in Eq. (\ref{errorestimation}).

\bigskip


\begin{thebibliography}{99}

\bibitem{Mukamel:Book} S. Mukamel, \textit{Principles of Nonlinear Optical Spectroscopy}, (Oxford University Press, USA 1999).

\bibitem{lit1} V.M. Kenkre, R.S. Knox, Phys. Rev. Lett. \textbf{33}, 803 (1974).

\bibitem{lit2} R. M. Pearlstein, J. Photochem. Photobiol. 35,Ê835 (1982).

\bibitem{lit3} V. M. Kenkre and P. Reineker, \textit{Exciton
Dynamics in Molecular Crystals and Aggregates} (Springer, Berlin, 1982).

\bibitem{lit4} S. Savikhin, D. R. Buck, W. S. Struve, The Fenna-Matthews- Olson protein: A strongly coupled photosynthetic antenna, in \textit{Resonance Energy Transfer}, Eds. D.C. Andrews and A.A. Demidov, (Wiley Interscience, New York 1998).

\bibitem{may1} Th. Renger and V. May, J. Phys. Chem. A \textbf{102}, 4381 (1998).


\bibitem{lit5} H. van Amerongen, L. Valkunas, R. van Grondelle, \textit{Photosynthetic excitons}, (Singapore World Scientific, 2000).

\bibitem{may2} B. Br\"{u}ggemann and V. May, J. Phys. Chem. B \textbf{108}, 10529 (2004).

\bibitem{lit6} V. May and O. Kuhn, \textit{Charge and Energy Transfer
Dynamics in Molecular Systems} (Wiley-VCH, Weinheim, 2004).


\bibitem{lit7} V. I. Novoderezhkin, M. A. Palacios, H. van Amerongen, R. van Grondelle, J. Phys. Chem. B \textbf{108}, 10363 (2004).

\bibitem{Adolph} J. Adolphs and T. Renger, Biophys. J., 91, 2778 (2006).

\bibitem{lit8} F. M\"{u}h, M. El-Amine Madjet, J. Adolphs, A. Abdurahman,
B. Rabenstein, H. Ishikita, E.-W. Knapp, and T.\ Renger, Proc. Natl. Acad.
Sci. USA \textbf{104}, 16862 (2007).


\bibitem{Engel07} G.S. Engel, T. R. Calhoun, E. L. Read, T. K. Ahn, T. Mancal, Y. C. Cheng, R. E. Blankenship and
G. R. Fleming, Nature 446, 782 (2007).

\bibitem{Lee07} H. Lee, Y.-C. Cheng, and G.R. Fleming, Science 316, 1462
(2007).

\bibitem{Calhoun} T. R. Calhoun, N. S. Ginsberg, G. S. Schlau-Cohen, Y.-C. Cheng, M.
Ballottari, R. Bassi, and G. R. Fleming, J. Phys. Chem. B 113, 16291
(2009).

\bibitem{Mercer09} I. Mercer, Y. El-Taha, N. Kajumba, J. Marangos, J. Tisch, M. Gabrielsen, R. Cogdell, E. Springate, and E. Turcu, Phys. Rev. Lett. 102,
057402 (2009).

\bibitem{Scholes09-1} E. Collini and G. D. Scholes, Science 323, 369 (2009).

\bibitem{Scholes09-2} E. Collini, C. Y. Wong, K. E. Wilk, P. M. Curmi, P. Brumer, and G. D.
Scholes, Nature 463, 644 (2010).

\bibitem{panit} G. Panitchayangkoon, D. Hayes, K. A. Fransted, J. R. Caram,
E. Harel, J. Wen, R. E. Blankenship, G. S. Engel, Proc. Nat. Acad. Sci 107, 12766 (2010).

\bibitem{mohseni-fmo} M. Mohseni, P. Rebentrost, S. Lloyd, and A. Aspuru-Guzik, J. Chem. Phys. 129, 174106
(2008).

\bibitem{Plenio08-1} M.B. Plenio and S.F. Huelga, New J. Phys. 10, 113019
(2008).

\bibitem{Castro08} A. Olaya-Castro, C. Fan Lee, F. Fassioli Olsen, and N. F.
Johnson, Phys. Rev. B \textbf{78}, 085115 (2008).

\bibitem{Rebentrost08-1} P. Rebentrost, M. Mohseni, A. Aspuru-Guzik,
J. Phys. Chem. B 113, 9942 (2009).

\bibitem{Rebentrost08-2} P. Rebentrost, M. Mohseni, I. Kassal, S. Lloyd, and A. Aspuru-Guzik,
New J. of Phys., 11, 033003 (2009).

\bibitem{Plenio09} F. Caruso, A. W. Chin, A. Datta, S. F. Huelga, M. B.
Plenio, J. of Chem. Phys. 131, 105106 (2009).

\bibitem{AkiPNAS} A. Ishizaki, G.R. Fleming, Proc. Nat. Acad. Sci USA 106, 17255 (2009).

\bibitem{Asadian} A. Asadian, M. Tiersch, G. G. Guerreschi, J. Cai,
S. Popescu, and H. J. Briegel, New. J. Phys. 12, 075019 (2010).

\bibitem{Aki-PCCP} A. Ishizaki, T. R. Calhoun, G. S. Schlau-Cohen and G. R. Fleming, Phys. Chem. Chem. Phys. 12, 7319 (2010).

\bibitem{Shim12} S. Shim, P. Rebentrost, S. Valleau, A. Aspuru-Guzik, Biophys. J. 102, 649 (2012).

\bibitem{CaoSilbey} J. Cao, R. Silbey, J. Phys. Chem. A 113, 13826 (2009).

\bibitem{Sarovar} M. Sarovar, A. Ishizaki, G. R. Fleming, and K. B. Whaley, Nature Physics 6, 462 (2010).

\bibitem{Fassioli} F. Fassioli and A. Olaya-Castro, New. J. Phys. 12, 085006 (2010).

\bibitem{Plenio10} F. Caruso, A. W. Chin, A. Datta, S. F. Huelga, M. B. Plenio, Phys. Rev. A 81, 062346 (2010).


\bibitem{masoud-tomography} J. Yuen-Zhou, M. Mohseni, A. Aspuru-Guzik, arXiv:1006.4866.

\bibitem{Xiong} J. Xiong and C. E. Bauer, Annu. Rev. Plant Biol. 53, 503 (2002).



\bibitem{Ishizaki09} A. Ishizaki and G. R. Fleming, J. Chem. Phys. 130,
234110 (2009).

\bibitem{H-S} H. Haken, G. Strobl, Z. Physik 262, 35 (1973).

\bibitem{H-S-1} P. Reineker, Springer Tracts in Modern Physics 94, 111 (Berlin, Heidelberg, New York: Springer-Verlag 1982).


\bibitem{Ishizaki09-2} A. Ishizaki and G. R. Fleming, J. Chem. Phys. 130,
234111 (2009).

\bibitem{Tanimura} R. Kubo, Adv. Chem. Phys. 15, 101 (1969).
\bibitem{Tanimura-1} Y. Tanimura and R. Kubo, J. Phys. Soc. Jpn. 58, 101 (1989).
\bibitem{Tanimura2} Y. Tanimura, J. Phys. Soc. Jpn. 75, 082001 (2006).
\bibitem{Cao} J. Cao, J. Chem. Phys. 107, 8 (1997).
\bibitem{N1} S. Jang, J. Cao, and R. J. Silbey,
J. Chem. Phys. 116, 2705 (2002); 
\bibitem{N2}S. Jang, M.D. Newton, and R.J. Silbey, Phys. Rev. Lett. \textbf{92}, 218301 (2004).

\bibitem{Jang07} S. Jang, M.D. Newton, and R.J. Silbey,
J. Phys. Chem. B \textbf(111), 6807 (2007).
\bibitem{N3} S. Jang, Y. C. Cheng, D. R. Reichman, and J. D. Eaves, J. Chem. Phys. 129, 101104 (2008).
\bibitem{Patrick3} P. Rebentrost, R. Chakraborty, A. Aspuru-Guzik, J. Chem. Phys. 131, 184102 (2009).
\bibitem{N4}A. Nazir, Phys. Rev. Lett. 103, 146404 (2009).
\bibitem{N5} J. Roden, A. Eisfeld, W. Wolff, and W. Strunz, Phys. Rev. Lett. 103, 058301 (2009).

\bibitem{Shi} Q. Shi, L. Chen, G. Nan, R. X. Xu, and Y. J. Yan, J. Chem. Phys. 130, 084105 (2009).
\bibitem{N6} G. Tao and W. H. Miller, J. Phys. Chem. Lett. 1, 891 (2010).
\bibitem{N7} J. Prior, A. W. Chin, S. F. Huelga, and M. B. Plenio, Phys. Rev. Lett. 105, 050404 (2010).
\bibitem{N8} X. T. Liang, Phys. Rev. E 82, 051918 (2010); 
\bibitem{N8-1} P. Nalbach and M. Thorwart, J. Chem. Phys. 132, 194111 (2010).
\bibitem{N9} I. D. Vega, arXiv:1005.0465.
\bibitem{N10} H. Fujisaki, Y. Zhang, and J. E. Straub, arXiv:1003.4796.
\bibitem{Bourret-1} X. Chen and R. J. Silbey, J. Phys. Chem. B 115, 5499 (2011).
\bibitem{Ritschel} G. Ritschel, J. Roden, W. T. Strunz, A. Eisfeld, New J. Phys. 13, 113034 (2011).

\bibitem{comp} M. Mohseni, A. Shabani, S. Lloyd and H. Rabitz, arXiv:1104.4812.

\bibitem{thebook} \textit{Quantum Effects in Biology} Edited by M. Mohseni,
Y. Omar, G. Engel, and M. Plenio, in preparation (Cabmeridge
University Press, Cambridge, UK, 2011).

\bibitem{Cho} M. Cho, H. M. Vaswani, T. Brixner, J. Stenger, and G. R. Fleming, J. Phys. Chem. B. 109, 10542 (2005).

\bibitem{Prall} B. S. Prall, D. Y. Parkinson, M. Yang, N. Ishikawa, G. R. Fleming, J. Chem. Phys. 120, 2537 (2004).

\bibitem{gWick1} C. M. Van Vliet, \textit{Equilibrium and non-equilibrium statistical mechanics},
(World Scientific, Singapore, 2008).

\bibitem{gWick2} K. Goldstein, D. A. Lowe, Nucl. Phys. B 669, 325 (2003).

\bibitem{Shabani} A. Shabani and D. Lidar, Phys. Rev. A, 80, 012309 (2009).

\bibitem{Rugh} W. Rugh, \textit{Linear System Theory},
(2nd ed. ~Prentice-Hall, Upper Saddle River, NJ, 1996).

\bibitem{DFT} R. G. Parr, W. Yang, \textit{Density-functional theory of atoms and molecules},
(Oxford University Press, 1994).

\bibitem{DMRG} H. Fehske, R. Schneider, and A. Wei§e (Eds.),
\textit{Computational Many Particle Physics}, Lecture Notes in Physics 739, Springer-Verlag, Berlin, Heidelberg (2008).

\bibitem{Breuer} H. -P. Breuer and F. Petruccione, \textit{The Theory of
Open Quantum Systems} (Oxford Univerity Press, New York, 2002).

\bibitem{Bourret} D. R. Reichman, F. L. H. Brown, and P. Neu, Phys. Rev. E 55, 2328 (1997).


\bibitem{Read1} D. Zigmantas, E. L. Read, T. Mancaÿl, T. Brixner, A. T. Gardiner, R. J.
Cogdell, and G. R. Fleming, Proc. Natl. Acad. Sci. USA 103, 12672 (2006).

\bibitem{Read2} EL Read, G. S. Engel, T. R. Calhoun, T. Mancaÿl, T. K. Ahn, R. E.
Blankenship, and G. R. Fleming, 104, 14203 (2007).

\bibitem{Read3} EL Read, G. S. Schlau-Cohen, G. S. Engel, J. Wen, R. E. Blankenship,
and G. R. Fleming, Biophys J 95, 847 (2008).

\bibitem{Brixner} T. Brixner, J. Stenger, H. M. Vaswani, M. Cho, R. E. Blankenship, and G.
R. Fleming, Nature 434, 625 (2005).

\bibitem{rec} T.G. Owens, S.P. Webb, L. Mets, R.S. Alberte, and G.R. Fleming,, Proc. Natl. Acad. Sci. USA 84, 1532 (1987).


\bibitem{Patrick2} P. Rebentrost and A. Aspuru-Guzik, J. Chem. Phys. 134, 101103 (2011).

\bibitem{Darius} D. Chruscinski, A. Kossakowski, S. Pascazio, Phys. Rev. A 81, 032101 (2010).

\bibitem{Ritz} T. Ritz, S. Park, and K. Schulten, J. Phys. Chem. B 105, 8259 (2001).

\bibitem{McKenzie} J. Gilmore and R. H. McKenzie, J. Phys. Chem. A, 112, 2162 (2008).

\bibitem{McKenzie-1} A. Damjanovi\v{c}, I. Kosztin, U. Kleinekath\"{o}fer, and K. Schulten, Phys. Rev. E 65, 031919 (2002).


\bibitem{Wu} J. Wu, F. Liu, Y. Shen, J. Cao and R. J Silbey,
New. J. Phys. 12, 105012 (2010).

\bibitem{modified} M. Yang and G. R. Fleming, Chem Phys 275, 355 (2002).

\bibitem{filtered} Q. Shi, L. Chen, G. Nan, R. X. Xu, and Y. J. Yan; J. Zhu, S. Kais, P. Rebentrost and A. Aspuru-Guzik, J. Phys. Chem. B 115, 1531-1537 (2011). 

\bibitem{Ng2} H. K. Ng, D. A. Lidar, and J. Preskill, Phys. Rev. A 84, 012305 (2011).

\bibitem{Ng} H. K. Ng and J. Preskill, Phys. Rev. A 79, 032318 (2009).

\bibitem{AkiLowT} A. Ishizaki and Y. Tanimura, J. Phys. Soc. Jpn. 74, 3131 (2005).

\end{thebibliography}
\end{document}